\documentclass[fp,twocolumn]{jpsj3}
\recdate{\today}
\usepackage{txfonts}
\usepackage{bm}
\usepackage{subfigure}

\newcommand{\brac}[1]{\left[ #1 \right]}
\newcommand{\brkt}[3]{\langle #1 | #2 | #3 \rangle}

\newcommand{\bk}{\bm{k}}

\newcommand{\Fenav}{\mathcal{F}_{\mathrm{SPA}}}
\newcommand{\Ffirstav}{\mathcal{F}_{\mathrm{SPA}}^{\,\mathrm{I}}}

\newcommand{\Q}[1]{\mathcal{Q}_{ #1 }}
\newcommand{\crpf}[1]{f^{\mathrm{phys}\dagger}_{#1}}
\newcommand{\anpf}[1]{f^{\mathrm{phys}}_{ #1 }}

\newcommand{\crf}[1]{f^\dagger_{#1}}
\newcommand{\anf}[1]{f_{ #1 }}

\newcommand{\crb}[1]{\phi^\dagger_{ #1 }}
\newcommand{\anb}[1]{\phi_{ #1 }}
\newcommand{\crbav}[1]{\overline{\phi}^{\,\ast}_{ #1 }}
\newcommand{\anbav}[1]{\overline{\phi}_{ #1 }}

\newcommand{\lamdav}[1]{\overline{\lambda}_{ #1 }}
\newcommand{\nlat}{N_L}

\newcommand{\II}{I\hspace{-.1em}I\hspace{+.1em}}
\newcommand{\III}{I\hspace{-.1em}I\hspace{-.1em}I\hspace{+.1em}}

\title{Slave Boson Analysis on $f^2$-Configuration System with $\Gamma_1$ Singlet Crystalline-Electric-Field Ground State in Cubic Symmetry}
\author{Taichi Hinokihara\thanks{hinokihara@spin.phys.s.u-tokyo.ac.jp}}
\inst{Department of Physics, Faculity of Science, The University of Tokyo, 7-3-1, Hongo, Bunkyo-ku, Tokyo 113-0033, Japan
}

\abst{ 
	We investigated electron phases in the $f^2$-configuration system with the $\Gamma_1$ crystalline-electric-field (CEF) ground state in cubic symmetry.
	An extended three-orbital periodic Anderson model, which consists of $j=5/2$ states, is evaluated by using the rotationally invariant slave boson formalism.
	As a result, we found three phases at zero temperature: the $\Gamma_1$ CEF singlet phase, and two kinds of Fermi liquid (FL).
	One of the FL arises owing to the presence of the $\Gamma_5$ CEF state, while the other forms a quasi-degenerated CEF state.
	All these phases are separated by first-order transitions: the charge-transfer transition and the CEF energy-level crossing.
	We also found heavy quasiparticle behaviors in the intermediate valence region between the $f^2$ and $f^3$ configurations.
	In this region, the mass enhancement factor exhibits a peak structure by increasing the hybridization.
	Physical properties of these phases and transitions are discussed.
}

\begin{document}
\maketitle
\section{Introduction}

Heavy electron systems, such as Ce-based, Pr-based, and U-based compounds, have been studied intensively because of its rich physical properties in the consequence of strongly correlated effects~\cite{Stewart1984,Pfleiderer2009}.
In the case of Ce-based compounds whose $f$-electron valency is around one ($f^1$-configuration system),
analysis on the single-orbital periodic Anderson model reveals basic properties of these systems despite the $f$-orbital degeneracy ($l=3$ and $s=1/2$) \cite{Rice1985,Yamada1986}.
The success of this approach is supported by the following two reasons:
a crystalline-electric-field (CEF) effect and the spin-orbit coupling split the $f$-electron degeneracy into several Kramers multiplets;
CEF excited states (ES) give essentially no influence on the low energy properties.
Thus, when the CEF ground state (GS) is a Kramers doublet, the single-orbital periodic Anderson model becomes an effective model for $f^1$-configuration systems.
In this situation, the same $f$-electron configurations are obtained between CEF GS and the quasiparticle (QP).

On the contrary, in the case of U-based and Pr-based compounds whose $f$-electron valency is around $f^2$ ($f^2$-configuration system), such the correspondence is no longer held.
The CEF states in the $f^2$-configuration can be written in the linear combination of Fock states (see Table \ref{tab:1}), while the QP is not necessarily in accordance with this configuration.
For this reason, it is questionable whether approximations based on the itinerant $f$-electron nature, such as the perturbation theory against electron-electron interactions, can treat the properties originated from the CEF states.
Likewise, it is also questionable whether approximations based on the localized $f$-electron nature, such as the Kondo lattice model, can reveal the QP properties.
Therefore, approximations treating both the itinerant and localized $f$-electron nature are required to investigate $f^2$-configuration systems.

Slave boson formalisms are the simplest approach satisfying this requirement.
Reference ~\citen{Lechermann2007} has developed the rotationally invariant slave boson (RISB) formalism, which can apply to general multi-orbital strongly-correlated-electron systems.
The RISB saddle point approximation (SPA) can evaluate the characteristic properties of QP, such as the renormalization factor and the QP energy level.
Moreover, we can also evaluate the localized $f$-electron behaviors, i.e., the expectation values of CEF states.
In particular, in the atomic limit, the RISB formalism can reproduce the correct CEF energy level scheme.

In this paper, we focus on the $f^2$-configuration system with the $\Gamma_1$ singlet CEF GS in cubic symmetry.
The filled skutterudite compound $\mathrm{PrOs_4Sb_{12}}$ belongs to this system and exhibits the heavy QP behavior~\cite{Bauer2002,Tayama2003,Kuwahara2004}.
In addition, several studies have suggested that physical properties observed in $\mathrm{UBe_{13}}$ can be explained by the $\Gamma_1$ CEF GS system~\cite{Kim1990,Nishiyama2013phD}.
$\mathrm{UBe_{13}}$ exhibits large mass enhancement and the unconventional non Fermi liquid (NFL)~\cite{Ott1983}.
Hence, how the $\Gamma_1$ CEF GS relates to these rich physical properties are need to be confirmed.

To investigate the $\Gamma_1$ CEF GS system, previous studies have employed the singlet-triplet model, which contains the $\Gamma_7$ doublet and $\Gamma_8$ quartet CEF states in the $f^1$-configuration, and the $\Gamma_1$ singlet CEF GS and the $\Gamma_4$ triplet CEF first ES in the $f^2$-configuration \cite{Hattori2005,Nishiyama2013phD}.
However, it has been not confirmed whether the excluded CEF states in the singlet-triplet model are surely ineffective.
As an example, treating CEF states in the $f^3$-configuration may play an important role for the itinerant properties of the $f$-electrons around the $f^2$-configuration.
Thus, we employ the three-orbital Anderson model consisting of the $j=5/2$ in cubic symmetry, which contains all the CEF states up to the $f^6$-configuration.
We tune the CEF parameter so as to be the $\Gamma_1$ CEF GS system and investigate this model by using the RISB SPA.

\begin{figure*}[t]
	\centering
	\subfigure[]{
		\includegraphics[width=0.4\linewidth]{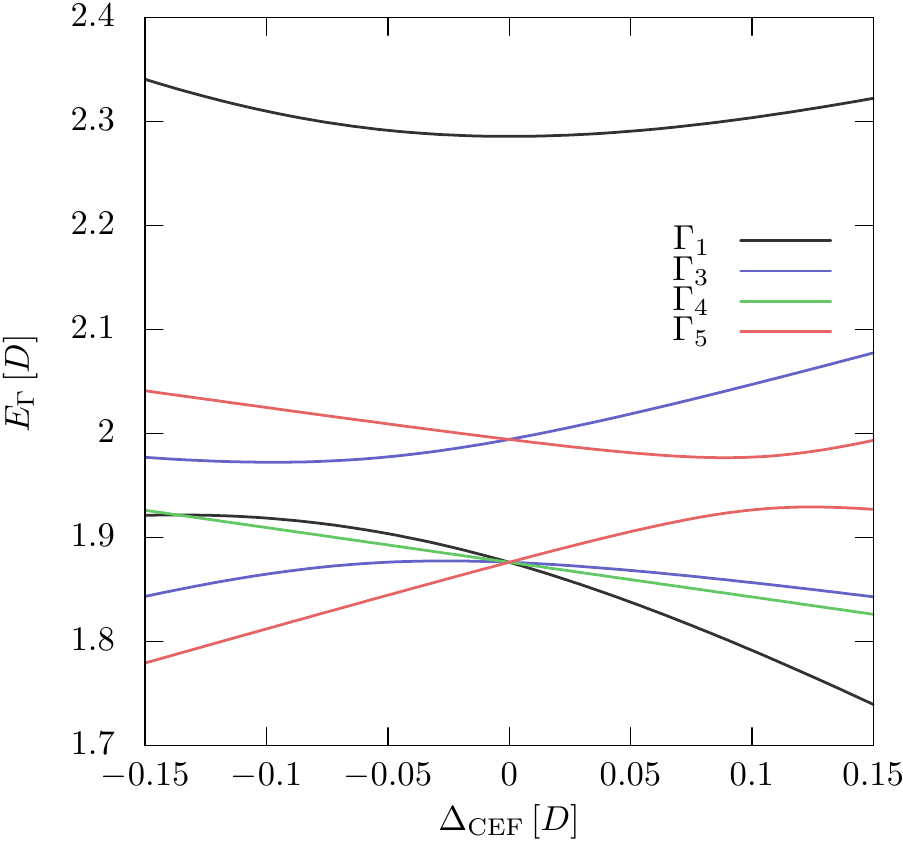}
	\label{Fig:cef2}
	}
	\subfigure[]{
	\includegraphics[width=0.4\linewidth]{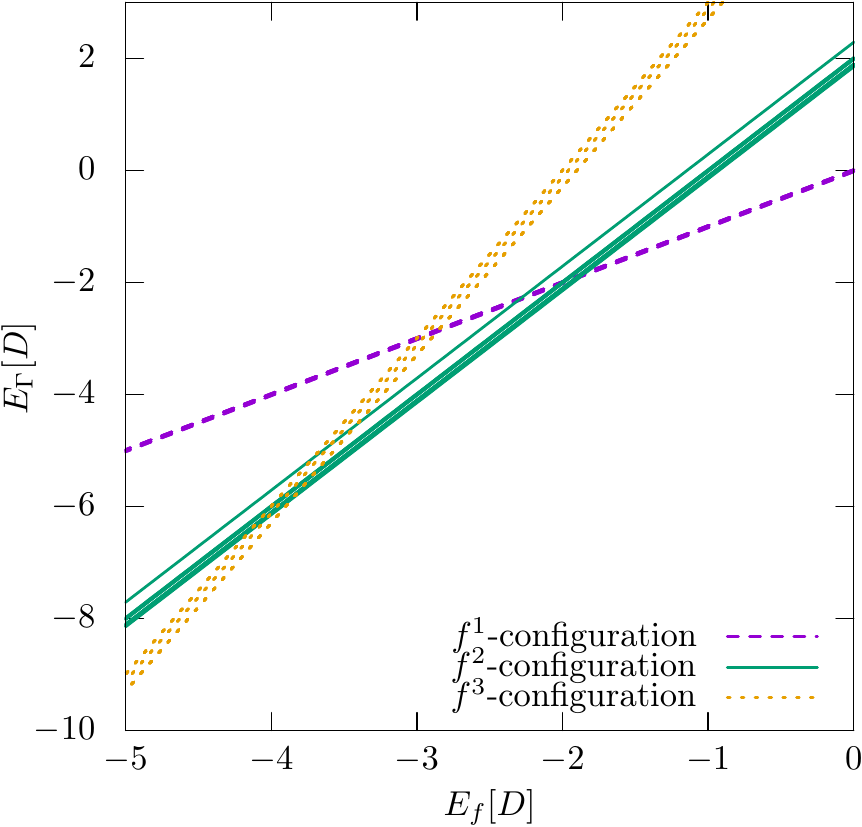}
	\label{Fig:Ef123}
	}
	\caption{(a) CEF energy level scheme of the $f^2$-configuration, in the $U=2.0D$ case, as a function of energy gap $\Delta_{\mathrm{CEF}}$.
	Number in parenthesis denotes the degeneracy of the CEF state.
		(b)CEF energy levels $E_{\Gamma}$ up to the $f^3$-configuration as a function of the $f$-electron energy level $E_f -\mu$.}
\end{figure*}

The organization of this paper is as follows.
In Sect. 2, we introduce the effective three-orbital periodic Anderson model and discuss the CEF energy-level scheme in the $f^2$-configuration.
In Sect. 3, we briefly introduce the RISB SPA.
In Sect. 4, the phase diagram of the $\Gamma_1$ CEF GS system is exhibited by using the RISB SPA.
Properties of phases and transitions are also discussed.
In Sect. 5, different points of the phase diagram from the singlet-triplet model is discussed.
Finally, we summarize this paper in Sect. 6.

\section{Effective Hamiltonian for $f$-electron systems}

\begin{table*}[t!]
	\centering
	\begin{tabular}{c|c}
		\shortstack{Irreducible\\ Representation} & Eigenstate in Terms of $f^1$-configuration State \\\hline
		$\Gamma_1$    & $-0.8588\left\{\Gamma_{+7},\Gamma_{-7 }\right\} +0.3622\left\{\Gamma_{+81},\Gamma_{-81}\right\} +0.3622\left\{\Gamma_{+82},\Gamma_{-82}\right\} $\\
		&\\
		$\Gamma_4(1)$ & $ 0.7071\left\{\Gamma_{+7},\Gamma_{-81}\right\} +0.7071\left\{\Gamma_{-7},\Gamma_{+81}\right\} $\\
		$\Gamma_4(2)$ & $-0.5000\left\{\Gamma_{+7},\Gamma_{+81}\right\} +0.8660\left\{\Gamma_{-7},\Gamma_{-82}\right\} $\\
		$\Gamma_4(3)$ & $0.8660\left\{\Gamma_{+7},\Gamma_{+82}\right\}  -0.5000\left\{\Gamma_{-7},\Gamma_{-81}\right\} $\\
		&\\
		$\Gamma_3(1)$ & $0.6513\left\{\Gamma_{+7},\Gamma_{-81}\right\} -0.6513\left\{\Gamma_{-7},\Gamma_{-81}\right\} -0.2750\left\{\Gamma_{+81},\Gamma_{-81}\right\}+0.2750\left\{\Gamma_{+82},\Gamma_{-82}\right\} $\\
		$\Gamma_3(2)$ & $0.6513\left\{\Gamma_{+7},\Gamma_{-82}\right\} -0.6513\left\{\Gamma_{-7},\Gamma_{+82}\right\} +0.2750\left\{\Gamma_{+81},\Gamma_{-82}\right\}-0.2750\left\{\Gamma_{-81},\Gamma_{+82}\right\}$\\
		&\\
		$\Gamma_5(1)$ & $ 0.2631\left\{\Gamma_{+7},\Gamma_{+81}\right\} +0.1519\left\{\Gamma_{-7},\Gamma_{-82}\right\} +0.9527\left\{\Gamma_{-81},\Gamma_{-82}\right\}$\\
		$\Gamma_5(2)$ & $-0.2148\left\{\Gamma_{+7},\Gamma_{-82}\right\}-0.2148\left\{\Gamma_{-7},\Gamma_{+82}\right\}+0.6737\left\{\Gamma_{+81},\Gamma_{-82}\right\}+0.6737\left\{\Gamma_{-81},\Gamma_{+82}\right\}$\\
		$\Gamma_5(3)$ & $ 0.1519\left\{\Gamma_{+7},\Gamma_{+82}\right\}+0.2631\left\{\Gamma_{-7},\Gamma_{-81}\right\}+0.9527\left\{\Gamma_{+81},\Gamma_{+82}\right\}$\\
	\end{tabular}
	\caption{Eigenstates of CEF states up to third excited state in the $f^2$-configuration in the case of $B_{40}=0.0001D$ and $U=2.0D$.
		Braces on the right hand side are defined as $\left\{\nu_1,\nu_2 \right\} = \frac{1}{\sqrt{2}}\left(|\nu_1,\nu_2\rangle - |\nu_2,\nu_1\rangle\right)$.}
	\label{tab:1}
\end{table*}

Let us introduce the effective Hamiltonian $\mathcal{H}$ that can represent the $\Gamma_1$ CEF GS system in cubic symmetry.
In this paper, we employ the three-orbital periodic Anderson model composed of $j=5/2$ states with the CEF splitting.
We first consider the atomic limit and introduce the exact localized $f$-electron Hamiltonian $\mathcal{H}_{\mathrm{loc}}$ for $j=5/2$ states.
Then, we introduce the itinerant Hamiltonian $\mathcal{H}_{\mathrm{itin}}$, which consists of the energy dispersion of the conduction electrons and the hybridizations between the conduction and $f$-electrons.

In the atomic limit, the fourteen-fold degeneracy of the $f$-electron ($l=3$ and $s=1/2$) splits into several CEF multiplets by the spin-orbit coupling, the Coulomb interactions, and the CEF effect.
Owing to the large spin-orbit coupling on the $f$-orbital, we consider infinitely large spin-orbit coupling to reduce the number of orbitals in the effective Hamiltonian.
Namely, hereafter we focus on the three orbital system consisting of $j=5/2$ (3-1/2) states and ignore $j=7/2$ (3+1/2) states.
The eigenstates for the $j=5/2$ are given by
\begin{align}
	\begin{cases}
			|\pm5/2\rangle = \pm \sqrt{\frac{1}{7}}|\pm2,\uparrow\rangle \mp \sqrt{\frac{6}{7}}|\pm3,\downarrow\rangle,\\
			|\pm3/2\rangle = \pm \sqrt{\frac{2}{7}}|\pm1,\uparrow\rangle \mp \sqrt{\frac{5}{7}}|\pm2,\downarrow\rangle,\\
			|\pm1/2\rangle = \pm \sqrt{\frac{3}{7}}| 0,\uparrow\rangle \mp \sqrt{\frac{4}{7}}|\pm1,\downarrow\rangle,
	\end{cases}
	\label{eq_j52}
\end{align}
where the basis $|l_z,s_z\rangle$ on the right hand side (rhs) consists of the orbital angular momentum $l_z$ ($l=3$) and the spin angular momentum $s_z$ ($s=1/2$).

In cubic symmetry, the six-fold degeneracy is lifted by the CEF Hamiltonian $\mathcal{H}_{\mathrm{CEF}}$ written as
\begin{align}
	\mathcal{H}_{\mathrm{CEF}} &= B_{40} \left( \hat{O}_{40} + 5\hat{O}_{44}\right) + B_{60}\left( \hat{O}_{60} - 21\hat{O}_{64}\right),
	\label{CEF_cubic_arb}
\end{align}
where $B_{nm}$ and $\hat{O}_{nm}$ denote the CEF parameters and Stevens operators for the $j=5/2$, respectively~\cite{Hutchings1964}.
Thus, the one-body $f$-electron eigenstates $|\nu\rangle$ and their energies $\varepsilon_\nu$ are obtained as follows:
\begin{align}
\label{eq_eigeneng_cubic1}
	\varepsilon_{\Gamma_7} = -240B_{40}, \\
\label{eq_eigen_cubic1}
	   	    &|\Gamma_{\pm 7}\rangle = \sqrt{\frac{1}{6}}|\pm\frac{5}{2}\rangle - \sqrt{\frac{5}{6}}|\mp\frac{3}{2}\rangle,\\
\label{eq_eigeneng_cubic2}
	\varepsilon_{\Gamma_8} = 120B_{40},\\
	&
\label{eq_eigen_cubic2}
	\begin{cases}
			&|\Gamma_{\pm 81}\rangle=\sqrt{\frac{5}{6}}|\pm\frac{5}{2}\rangle + \sqrt{\frac{1}{6}}|\mp\frac{3}{2}\rangle,\\
			&|\Gamma_{\pm 82}\rangle=|\pm\frac{1}{2}\rangle,
	\end{cases}
\end{align}
where the basis $|\nu\rangle$ on the left hand side (lhs) are written in terms of the irreducible representation of point group in cubic symmetry.
Note that the sixth-order of Stevens operators give no influence on the $j=5/2$ states, and thus the only CEF parameter $B_{40}$ determines the CEF energy splitting.
The energy gap $\Delta_{\mathrm{CEF}}$ between the two states is equivalent to $360B_{40}$.

In the plural $f$-electron systems, the Coulomb interactions also affects CEF states.
Since we derived the diagonalized one-body terms as written in Eqs.~\eqref{eq_eigen_cubic1} and \eqref{eq_eigen_cubic2}, it is convenient to express the Coulomb interactions $H_U$ in terms of these states.
According to Refs.~\citen{Slater1929,Condon1931,Racah1942}, the Coulomb interactions among $f$-electrons ($l=3$) $\mathcal{H}^{ff}_U$ can be described by using Slater-Condon parameters $F^k$ and Gaunt coefficients $c_k(m_z, m_z^\prime)$ as follows:
\begin{align}
	\mathcal{H}^{ff}_U = \sum_{m_{z1}\ldots m_{z4}}\sum_{\sigma_{z}}
	&I_{m_{z1}m_{z2}}^{m_{z3}m_{z4}}f^{\mathrm{phys}\dagger}_{m_{z4}\sigma_{z}} f^{\mathrm{phys}\dagger}_{m_{z3}\sigma_{z}}f^{\mathrm{phys}}_{m_{z2}\sigma_{z}} f^{\mathrm{phys}}_{m_{z1}\sigma_{z}}\nonumber\\
	+&I_{m_{z1}m_{z2}}^{m_{z3}m_{z4}}f^{\mathrm{phys}\dagger}_{m_{z4}\sigma_{z}} f^{\mathrm{phys}\dagger}_{m_{z3}\overline{\sigma}_{z}}f^{\mathrm{phys}}_{m_{z2}\overline{\sigma}_{z}} f^{\mathrm{phys}}_{m_{z1}\sigma_{z}},
	\label{chap2:interaction}\\
\end{align}
where,
\begin{align}
	I_{m_{z1}m_{z2}}^{m_{z3}m_{z4}} \equiv
	\sum_{k=0}^6 F^k& c_k\left(m_{z1},m_{z4}\right) c_k\left(m_{z2},m_{z3}\right)\delta_{m_{z1}+m_{z2},m_{z3}+m_{z4}}.
	\label{interaction_ff}
\end{align}
Here, $c_k(m_z, m_z^\prime)$ has the following relation: $c_k(m_z, m_z^\prime) = (-1)^{\left(m_z - m_z^\prime\right)}c_k(m_z^\prime, m_z)$.
Since all the Gaunt coefficients with odd-number $k$ are zero, $\mathcal{H}^{ff}_U$ contains four Slater-Condon parameters, $F^0$, $F^2$, $F^4$, and $F^6$.
By performing the unitary transformation, i.e., transforming $|m_z s_z\rangle$ bases into $|\nu\rangle$ bases, $\mathcal{H}_U$ is given by
\begin{align}
	\mathcal{H}_U = &\sum_{\nu_{1}\ldots \nu_{4}}
	I_{\nu_{1}\nu_{2}}^{\nu_{3}\nu_{4}}f^{\mathrm{phys}\dagger}_{\nu_4} f^{\mathrm{phys}\dagger}_{\nu_3}f^{\mathrm{phys}}_{\nu_{2}} f^{\mathrm{phys}}_{\nu_{1}},\nonumber\\
	I_{\nu_{1}\nu_{2}}^{\nu_{3}\nu_{4}} \equiv \sum_{m_{z1}\ldots m_{z4}}\sum_\sigma&
	U^{\nu_4}_{m_{z4}\sigma_z}U^{\nu_3}_{m_{z3}\sigma_z}U^{\nu_2\ast}_{m_{z2}\sigma_z}U^{\nu_1\ast}_{m_{z1}\sigma_z}
	I_{m_{z1}m_{z2}}^{m_{z3}m_{z4}}\nonumber\\
	&+U^{\nu_4}_{m_{z4}\sigma_z}U^{\nu_3}_{m_{z3}\overline{\sigma}_z}U^{\nu_2\ast}_{m_{z2}\overline{\sigma}_z}U^{\nu_1\ast}_{m_{z1}\sigma_z}
	I_{m_{z1}m_{z2}}^{m_{z3}m_{z4}},
	\label{interaction_i}
\end{align}
where $f^{\mathrm{phys}\dagger}_{\nu}$ ($f^{\mathrm{phys}}_{\nu}$) denotes the creation (annihilation) operators for the $f$-electrons with $\nu$ orbital
and $U^{\nu}_{m_z\sigma_z}$ indicates the element of the unitary transformation, which are derived from Eqs.~\eqref{eq_j52},~\eqref{eq_eigen_cubic1}, and~\eqref{eq_eigen_cubic2}.
Note that we denote the original $f$-electron creation/annihilation operators with the superscript \lq\lq phys" in order to distinguish from the pseudo fermion operators $f^{\dagger}_{\nu}$ introduced later in the RISB formalism.

Hereafter, we fix the ratio of the Slater-Condon parameters as follows:
\begin{align}
	F^0 = U, F^2 = 0.5U, F^4=0.3U,F^6 = 0.1U,
	\label{chap2:slater_condon}
\end{align}
where $U$ is a scaling parameter of the electron-electron interactions.
This ratio has been introduced in Ref.~\citen{Hotta2006} and confirmed that CEF eigenstates are insensitive to the ratio when $U$ is sufficiently larger than CEF splitting, $\Delta_{\mathrm{CEF}}$.

As a result, $\mathcal{H}_\mathrm{loc}$ is written as
\begin{align}
	\mathcal{H}_{\mathrm{loc}} = \sum_{\nu} \left(E_f + \varepsilon_{\nu} -\mu\right) f^{\mathrm{phys}\dagger}_\nu f^{\mathrm{phys}}_\nu + \mathcal{H}_U,
	\label{ham_loc}
\end{align}
where $E_f$ and $\mu$ denote the $f$-electron energy level and the Fermi energy, respectively.
We measure $E_f$ from the center of the conduction band introduced later.

The present model takes into account finite Coulomb interactions, which is crucial for holding the relation of all the CEF energy levels and eigenstates consistently.
In the conventional $jj$-coupling scheme, which assumes infinitely large Coulomb interactions, the relation between the CEF parameters for the $f^1$-configuration ($j=5/2$) and those for the $f^2$-configuration ($J=4$) is unclear.

Let us confirm that $\mathcal{H}_\mathrm{loc}$ realizes the $\Gamma_1$ singlet CEF GS in the $f^2$-configuration.
Figure~\ref{Fig:cef2} shows the CEF energy levels of the $f^2$-configuration as a function of the energy splitting $\Delta_{\mathrm{CEF}}$ for the case of $U=2.0D$ and $E_f-\mu=0$.
At $\Delta_{\mathrm{CEF}}=0$, the CEF energy level splitting is consistent with that in the $jj$-coupling scheme: the $J=4$ nonet, the $J=2$ quintet, and the $J=0$ singlet.
In the finite $\Delta_{\mathrm{CEF}}$, the $J=4$ states split into the $\Gamma_1$ singlet, $\Gamma_3$ doublet, $\Gamma_4$ triplet, and $\Gamma_5$ triplet CEF states.
In the case of $\Delta_{\mathrm{CEF}}>0$, the $\Gamma_1$ CEF state becomes the GS, otherwise the $\Gamma_5$ CEF state becomes the GS.
Note that both the $\Gamma_3$ CEF state and the $\Gamma_4$ CEF state cannot be the GS in the present system.

In the atomic limit, $E_f-\mu$ determines which $f^n$-configuration system is stable.
Figure \ref{Fig:Ef123} shows the $(E_f-\mu)$ dependence of the CEF energy levels up to the $f^3$-configuration with $U=2.0D$ and $B_{40} = 0.0001D$.
This figure indicates that the $f^2$-configuration system is realized in the region $-4.0D<E_f-\mu<-2.0D$.
Therefore, we conclude that the $\Gamma_1$ CEF GS system is realized in the region $-4.0D<E_f-\mu<-2.0D$ and $B_{40}>0$.

The CEF eigenstates up to the third excited states are listed in Table~\ref{tab:1} in the case of $B_{40}=0.0001D$ and $U=2.0D$.
The $\Gamma_1$ CEF state consists of the doubly occupied states on each orbital, and thus it seems difficult to form the heavy QPs on the $\Gamma_7$ orbital.
Likewise, the $\Gamma_4$ CEF state seems ineffective to construct the heavy QPs composed of the only $\Gamma_8$ orbitals.
Further discussion for the relation between the heavy QPs and the CEF states needs to introduce the hybridizations between the conduction and $f$-electrons and to analyze the strongly-correlated system.
We will later discuss this relation in Sect. 5.

\subsection{itinerant part $\mathcal{H}_{\mathrm{itin}}$}

This paper focuses on how the $\Gamma_1$ CEF GS affects physical properties.
Since introducing a realistic energy dispersion makes this point unclear, we introduce the simple form of $\mathcal{H}_{\mathrm{itin}}$ given by
\begin{align}
	\mathcal{H}_{\mathrm{itin}} = \sum_{ij\nu} \left(t_{ij\nu} -\mu\delta_{i,j}\right)c^\dagger_{i\nu}c_{j\nu}  + \sum_{i\nu}V_{\nu} c^\dagger_{i\nu} f^{\mathrm{phys}}_{i\nu} + \mathrm{h.c.}.
	\label{ham_iti}
\end{align}
Here, $c^{\dagger}_{i\nu}$ ($c_{i\nu}$) denotes the creation (annihilation) operators for conduction electrons, which are specified by labels $\nu$ same as $f$-electrons.
$t_{ij\nu}$ and $V_\nu$ indicate the hopping of the conduction electrons and the hybridization between the $f$-electrons and the conduction electrons within the same $\nu$ orbital, respectively.
Inter-orbital hopping terms and inter-orbital hybridizations are omitted in the present model.

We assume the rectangular form of the density of states (DOS) $\rho_{\varepsilon}$:
\begin{align}
	\label{dos}
	\rho_{\varepsilon}= 
	\begin{cases}
		\frac{1}{2D} &|\varepsilon| < D \\
		 0 & \mathrm{otherwise}
	\end{cases},
\end{align}
where $D$ is half of the bandwidth of the DOS.
Hereafter, we set $D$ as a unit of energy.

\begin{figure}[t]
	\centering
	\includegraphics[width=0.95\linewidth]{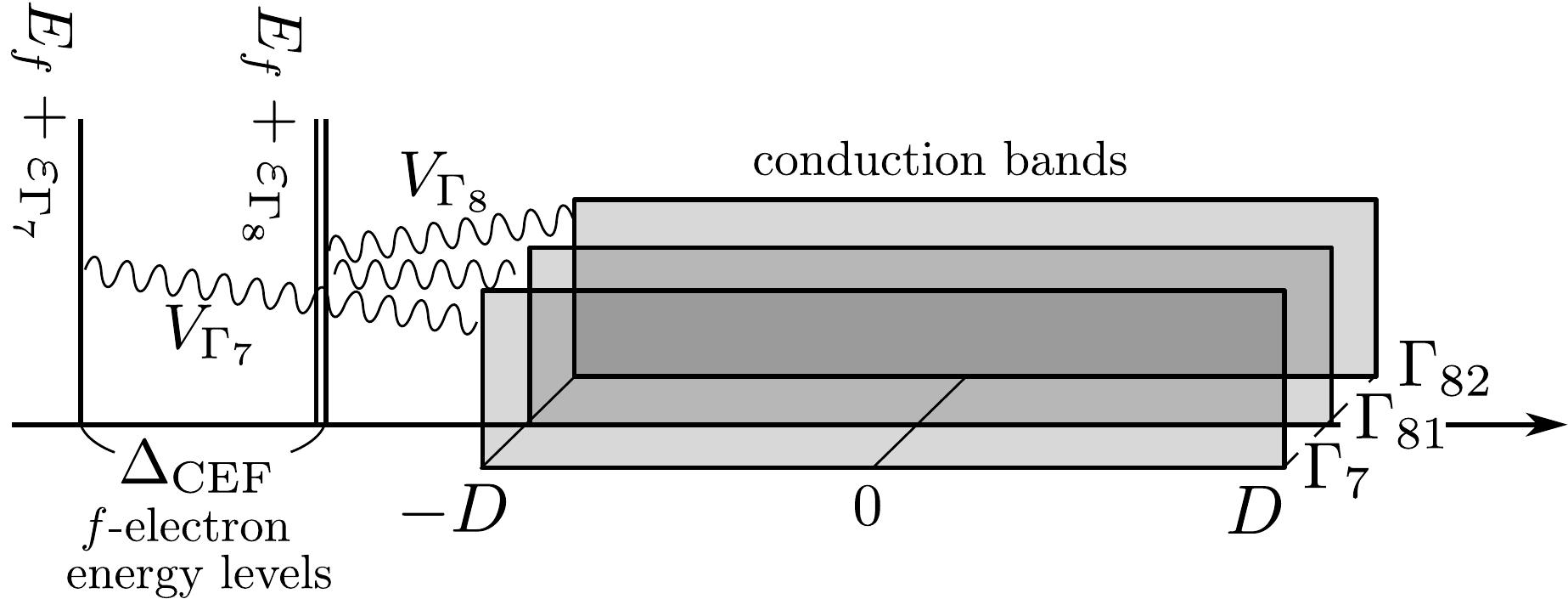}
	\caption{Schematic diagram of the present model.}
	\label{Fig:sch_dia}
\end{figure}

Finally, by combining $\mathcal{H}_{\mathrm{loc}}$ and $\mathcal{H}_{\mathrm{itin}}$, we obtain the effective Hamiltonian for the $f$-electron system as:
\begin{align}
	\label{org_h}
	\mathcal{H} =& \sum_{ij\nu} \left(t_{ij\nu} -\mu\delta_{i,j}\right)c^\dagger_{i\nu}c_{j\nu}  + \sum_{i\nu}\left(V_{\nu} c^\dagger_{i\nu} f^{\mathrm{phys}}_{i\nu} + \mathrm{h.c.}\right) \nonumber\\
	&+ \sum_{\nu} \left(E_f + \varepsilon_{\nu} -\mu\right) f^{\mathrm{phys}\dagger}_\nu f^{\mathrm{phys}}_\nu + \mathcal{H}_U.
\end{align}
The schematic picture of the present model is shown in Fig. \ref{Fig:sch_dia}.

\section{Rotationally invariant slave boson formalism}
In this section, we briefly introduce the RISB SPA in Ref.~\citen{Lechermann2007} and apply this formalism to Eq.~\eqref{org_h}.
The RISB formalism maps a physically meaningful state $|n\rangle$, which is a state in the original Hilbert space, into a state $|\underline{n}\rangle$ in the enlarged Hilbert space as follows:
\begin{align}
	\label{sq_risbst1}
	 |n\rangle \rightarrow|\underline{n}\rangle = \frac{1}{\sqrt{A_n}}\sum_{m}\crb{nm} |\mathrm{vac}\rangle_b \otimes | m\rangle_f.
\end{align}
Here, $A_n$ is the normalization factor that corresponds to a dimension of the subspace of the enlarged Hilbert space with particle number fixed as $n$.
$\crb{nm}$ ($\anb{nm}$) denotes the creation (annihilation) operator of slave bosons.
$|\mathrm{vac}\rangle_b$ denotes the vacuum state of the bosons.
$|m\rangle_f$ stands for the QP Fock state composed of pseudo fermions $f^\dagger_{\nu}$ as, 
\begin{align}
	|m\rangle_f = \prod_{\nu}\left(f^\dagger_{\nu}\right)^{m_\nu}|\mathrm{vac}\rangle_f,
\end{align}
where $m_\nu$ takes either 0 or 1.

In the RISB formalism, the boson operator $\crb{nm}$ is represented by pair of two different states, the physical state $n$ and the QP state $m$.
Since we do not discuss the superconductivity in this paper, the pairs of these states, $n$ and $m$, are restricted to have the same total particle number.
The RISB formalism ensures that results do not depend on the basis set, and thus we use the Fock state basis for both the physical states and the QP states.
Different basis, such as the eigenstate basis $\phi_{\Gamma m}$ and the Fock state basis $\phi_{n m}$, has the following relation:
\begin{align}
	\phi^\dagger_{\Gamma m} = \sum_m \langle n | \Gamma \rangle \phi^\dagger_{nm}.
	\label{eq_unit_boson}
\end{align}
We use this relation to evaluate the expectation values of the CEF states.

In order to exclude all the unphysical states, which are not included in the original Hilbert space but included in the enlarged Hilbert space,
the following constraint conditions are required:
\begin{align}
	\label{risb_constrainop1}
	\hat{\mathcal{Q}}_{0}|\underline{n}\rangle&= \left(\sum_{nm} \phi^\dagger_{nm}\phi_{nm} -1\right)|\underline{n}\rangle = 0,\\
	\label{risb_constrainop2}
	\hat{\mathcal{Q}}_{\nu\nu^\prime}|\underline{n}\rangle &= \left( f^\dagger_{\nu}f_{\nu^\prime}-\sum_{nml}\phi_{nm}^\dagger\phi_{nl}\langle l|\crf{\nu}\anf{\nu^\prime}|m\rangle_f\right)|\underline{n}\rangle = 0 .
\end{align}
We can easily check that all the physically meaningful states $|\underline{n}\rangle$ satisfy these conditions and all the other states do not.

By using the slave bosons and pseudo fermions, Eq. \eqref{org_h} is transformed as follows:
\begin{align}
	\label{sb_h}
	\underline{\mathcal{H}} &= \sum_{ij\nu} \left(t_{ij\nu} -\mu\delta_{i,j}\right)c^\dagger_{i\nu}c_{j\nu} 
	+ \sum_{i\nu}\left(\hat{R}_{\nu\nu^\prime}V_{\nu} c^\dagger_{i\nu} f_{i\nu^\prime} + \mathrm{h.c.}\right)\nonumber\\
	&+\sum_{inml} E_{nm} \phi^\dagger_{inl}\phi_{iml}.
\end{align}
where $E_{nm} = \langle n|\mathcal{H}_{\mathrm{loc}}|m\rangle$
and $\hat{R}_{\nu\nu^\prime}$ is the subsidiary operator consisting of the slave boson operators.
Although $\hat{R}_{\nu\nu^\prime}$ is not uniquely determined, it is confirmed that the following representation $\hat{R}_{\nu\nu^\prime}$ gives the same result of the Gutzwiller approximation:
\begin{align}
	\hat{R}_{\nu\nu^\prime} = 
	\sum_{\substack{n_1n_2 \\ m_1m_2 \\ \nu_1}} 
		\langle n_1|f^\mathrm{phys\dagger}_{\nu}|n_2\rangle {}_f\langle m_1 |f^\dagger_{\nu_1}|m_2 \rangle_f
		\hspace{-.15em}
	\phi^\dagger_{n_1m_1}\hat{M}^{m_2}_{\nu_1\nu^\prime}\phi_{n_2m_2},
	\label{subr}
\end{align}
where,
\begin{align}
	&\hat{M}^{m_2}_{\nu_1\nu^\prime} = 
	C^{\Q{0}-1}
	\left[
	\frac{1}{\sqrt{\hat{1} -\hat{\Delta}^{h}}}
	\frac{1}{\sqrt{\hat{1} -\hat{\Delta}^{p}}}
	\right]_{\nu_1\nu^\prime}
	,\\
	&C = \sqrt{\left(N_{m_2}+1\right)\left( 2N_{\mathrm{orb}} - N_{m_2}\right)}.
	\label{eq_M}
\end{align}
Here, $N_m$ denotes the particle number of the state $|m\rangle$ and $N_{\mathrm{orb}}$ denotes the number of orbitals, which corresponds to six in the present model.
$\hat{\Delta}^{p}$ ($\hat{\Delta}^{h}$) is the matrix of the particle (hole) operator whose $\nu\nu^\prime$ component is given by
\begin{align}
	\label{particle}
	\hat{\Delta}^{p}_{\nu\nu^\prime} = \sum_{nml}\phi_{nm}^\dagger\phi_{nl}\langle l|\crpf{\nu}\anpf{\nu^\prime}|m\rangle,\\
	\hat{\Delta}^{h}_{\nu\nu^\prime} = \sum_{nml}\phi_{nm}^\dagger\phi_{nl}\langle l|\anpf{\nu}\crpf{\nu^\prime}|m\rangle.
	\label{hole}
\end{align}

The SPA of the RISB regards all the slave bosons as mean values: $\phi_{nm} \rightarrow \overline{\phi}_{nm}$ and $\phi_{nm}^{\dagger} \rightarrow \overline{\phi}_{nm}^\ast$.
Since Eq. (\ref{org_h}) does not contain the inter-orbital hybridizations such as $V_{\Gamma_{+7}\Gamma_{+8}}$,
the particle (hole) operators $\hat{\Delta}^{p}_{\nu\nu^\prime}$ ($\hat{\Delta}^{h}_{\nu\nu^\prime}$) and the subsidiary operators $\hat{R}_{\nu\nu^\prime}$ are diagonalized in the SPA.
Thus, hereafter, we represents the mean values of these operators by $\overline{\Delta}^{p}_{\nu}$, and $\overline{R}_{\nu}$, respectively.

The free energy in the RISB SPA $\Fenav$ is given by
\begin{align}
	\Fenav =& \Ffirstav 
	     + N_L \sum_{n_1n_2m} E_{n_1n_2} \overline{\phi}_{n_1m}^\ast  \overline{\phi}_{n_2m}\nonumber  \\
		 & + N_L \sum_{nm} \left(\lambda_0 - \sum_{\nu}\langle m |f^\dagger_{\nu} f_{\nu} |m \rangle \lambda_{\nu}\right) |\overline{\phi}_{nm}|^2\nonumber\\
		 &- N_L\lambda_{0} + N\mu,
	\label{free_eng}
\end{align}
where $\lambda_{\nu}$ and $\lambda_{0}$ are the Lagrange multipliers of the constraint conditions [Eqs.~\eqref{risb_constrainop1} and \eqref{risb_constrainop2}].
$\Ffirstav$ is written as
\begin{align}
	\Ffirstav =&-\frac{1}{\beta}\sum_{\bk\sigma=\pm}\sum_{\nu} \ln \brac{ 1 + \exp\left[-\beta\right(\mathcal{E}^{\sigma}_{\bk\nu}-\mu\left)\right] }.
	\label{chap_g1:free_energy1st}
\end{align}
Here, $\mathcal{E}^{+}_{\bk\nu}$ ($\mathcal{E}^{-}_{\bk\nu}$) denotes the upper (lower) band energy dispersion given by
\begin{align}
	\mathcal{E}^{\pm}_{\bk\nu} = 
	\frac{1}{2} \left( \varepsilon_{\bk} + \lambda_\nu  \pm \sqrt{\left(\varepsilon_{\bk} - \lambda_\nu\right)^2 + 4z_{\nu}V^2_{\nu}} \right),
	\label{chap_g1:eigenenergy}
\end{align}
where $\varepsilon_{\bk}$ and $z_\nu$ denotes the energy dispersion of the conduction electrons and the renormalization factor $z_\nu = \overline{R}_\nu^2$ of the $\nu$ orbital, respectively.

The Lagrange multipliers, the chemical potential, and the mean values of all the slave bosons are determined by solving the following non-linear simultaneous equations:
\begin{align}
	\label{eqs1}
	\frac{1}{\nlat}\frac{\partial\Fenav}{\partial \lamdav{0}} =& \sum_{nm } \crbav{nm}\anbav{nm} - 1 = 0,
	\\
	\label{eqs2}
	\frac{1}{\nlat}\frac{\partial\Fenav}{\partial \lamdav{\nu}} =&
	\frac{1}{\nlat}\frac{\partial\Ffirstav}{\partial \lamdav{\nu}} - \sum_{nm} \brkt{m}{\crf{\nu}\anf{\nu}}{m}\crbav{nm}\anbav{nm} = 0,
	\\
	\label{eqs3}
	\frac{1}{\nlat}\frac{\partial\Fenav}{\partial \anbav{nm}} = &
	\frac{1}{\nlat}\frac{\partial\Ffirstav}{\partial \anbav{nm}} + \sum_{n^\prime} E_{n^\prime n}\crbav{n^\prime m} = 0,
	\\
	\label{eqs4}
	\frac{1}{\nlat}\frac{\partial\Fenav}{\partial \crbav{nm}} = &
	\frac{1}{\nlat}\frac{\partial\Ffirstav}{\partial \crbav{nm}} + \sum_{n^\prime} E_{n n^\prime}\anbav{n^\prime m} = 0,
	\\
	\label{eqs5}
	\frac{1}{\nlat}\frac{\partial\Fenav}{\partial \mu} = &
	\frac{1}{\nlat}\frac{\partial\Ffirstav}{\partial \mu} + N = 0.
\end{align}
Although we can solve these equations at finite temperature in principle, it is known that the artificial phase transition owing to the Bose-Einstein condensation (BEC) occurs at finite temperature.
Thus, we investigate this system at zero temperature.
The derivative of $\Ffirstav$ with respect to the mean-fields $\overline{\lambda}_{\nu}$, $\overline{\phi}_{nm}$, $\overline{\phi}_{nm}^\ast$, and $\mu$
can be analytically given by assuming the rectangular form of the DOS [Eq.~\eqref{dos}]:
\begin{align}
	\frac{1}{N_L}\frac{\partial\Ffirstav}{\partial \lambda_\nu} =& \sum_{\nu^\prime\sigma}\rho_0\int_{-D}^{D} d\varepsilon f(\mathcal{E}^{\sigma}_{\nu^\prime}\left(\varepsilon\right))\frac{\partial \mathcal{E}^{\sigma}_{\nu^\prime}\left(\varepsilon\right)}{\partial\lambda_{\nu}}\nonumber\\
	=& \sum_\sigma C^{\lambda}_{\nu\sigma},\\
	C^{\lambda}_{\nu\sigma} =&
	\begin{cases}
	\frac{\rho_0 z_{\nu}V_{\nu}^2}{\lambda_\nu - \min\left[\mu,\mathcal{E}^\sigma_{\nu}\left(D\right)\right]  } -\frac{\rho_0 z_{\nu}V_{\nu}^2}{\lambda_\nu - \mathcal{E}^\sigma_{\nu}\left(-D\right)} 
	& (\mu\geq \mathcal{E}^\sigma_{\nu}\left(-D\right))
	\\
	0
	& (\mu < \mathcal{E}^\sigma_{\nu}\left(-D\right)),
\end{cases}
	\label{int:f}
	\\
	\frac{1}{N_L}\frac{\partial\Ffirstav}{\partial \overline{\phi}_{nm}} =&  \sum_{\nu\sigma}\frac{\partial z_{\nu}}{\partial \overline{\phi}_{nm}}\rho_0\int_{-D}^{D} d\varepsilon f(\mathcal{E}^{\sigma}_\nu\left(\varepsilon\right))\frac{\partial \mathcal{E}^{\sigma}_\nu\left(\varepsilon\right)}{\partial z_{\nu}}\nonumber\\
	=& \sum_{\nu\sigma} C^{\phi}_{\nu\sigma},\\
	C^{\phi}_{\nu\sigma} =&
	\begin{cases}
		\rho_0\frac{\partial z_{\nu}}{\partial \overline{\phi}_{nm}}V_{\nu}^2 \ln|\frac{\lambda_{\nu} - \min\left[\mu,\mathcal{E}^\sigma_{\nu}\left(D\right)\right] }{\lambda_{\nu} - \mathcal{E}^\sigma_{\nu}\left(-D\right)}|
	& (\mu\geq \mathcal{E}^\sigma_{\nu}\left(-D\right))
	\\
	0
	& (\mu < \mathcal{E}^\sigma_{\nu}\left(-D\right)).
\end{cases}
	\label{int:q}
\end{align}
We numerically solve Eqs.~(\ref{eqs1})-(\ref{eqs5}) by using Broyden's method \cite{Broyden1965}.

Concluding this section, let us mention what we can discuss from the RISB SPA.
First, the pseudo fermions behaves as the QPs characterized by the QP energy level $\lambda_\nu$ and the renormalized hybridization $\sqrt{z_\nu}V$. 
Since the degenerated orbitals take the same $\lambda_\nu$ and $z_{\nu}$, hereafter we omit the subscript for the Kramers degeneracy, e.g., $\lambda_{\Gamma_7}\equiv\lambda_{\pm\Gamma_7}$ and $\lambda_{\Gamma_8}\equiv\lambda_{\pm\Gamma_{81}}=\lambda_{\pm\Gamma_{82}}$.
We also introduce the summation of the number of pseudo fermions over the degenerated orbitals as follows: 
\begin{align}
	\label{nG7}
	n_{\Gamma_7} &\equiv n_{\Gamma_{+7}} + n_{\Gamma_{-7}},\\
	\label{nG8}
	n_{\Gamma_8} &\equiv n_{\Gamma_{+81}} + n_{\Gamma_{-81}} + n_{\Gamma_{+82}} + n_{\Gamma_{-82}}.
\end{align}

According to Eq.~\eqref{eq_unit_boson}, we can also evaluate the expectation value of the CEF state $\Phi_{\Gamma}$.
Here, $\Phi_{\Gamma}$ is given by
\begin{align}
	\Phi_{\Gamma} &= \sum_{l} \langle \phi^\dagger_{\Gamma l} \phi_{\Gamma l}\rangle_{\mathrm{SPA}} = 
	\sum_{l} |\overline{\phi}_{\Gamma l}|^2 = \sum_{nml} \langle n |\Gamma\rangle\langle \Gamma |m\rangle\overline{\phi}_{n l}^{\ast} \overline{\phi}_{m l}.
\end{align}
By using $\Phi_{\Gamma}$,
the mean of the $f$-electron number $n_f$ and its variance $\sigma^2_f$ can be evaluated as follows:
\begin{align}
	n_f &= \sum_{\Gamma}N_{\Gamma} \Phi_{\Gamma}, \\
	\sigma_f^2 &= \sum_{N=1}^6\left( N \sum_{\Gamma}\delta_{N_{\Gamma},N}\Phi_{\Gamma}\right)^2  - n_f^2,
\end{align}
where $N_{\Gamma}$ denotes the particle number of the $\Gamma$ state.
We note that $n_f$ is equivalent to $\sum_\nu n_{\nu}$ owing to the constraint condition [Eq. \eqref{risb_constrainop2}].
The condition $\sigma_f/n_f\ll 1$ indicates that expectation values are almost exhausted by the CEF states in a $f^n$-configuration system, otherwise expectation values widely distribute to several $f^n$-configurations.
\section{Results}

Let us investigate the Hamiltonian [Eq. (\ref{org_h})] around the $f^2$-configuration by using the RISB SPA.
Parameters are fixed as $E_f=-3.3D$, $B_{40}=0.0001D$ and $U=2.0D$ so as to realize the $\Gamma_1$ CEF GS in the atomic limit as discussed in Sect. 2.
We also fix the total number of electrons per site $N$ as 4.0.
We evaluate this system in the region of $0.01D<V_{\Gamma_7}<0.75D$ and $ 0.01D<V_{\Gamma_8}<0.75D$.

\begin{figure*}[t!]
	\centering
	\subfigure[]{
		\includegraphics[width=0.4\linewidth]{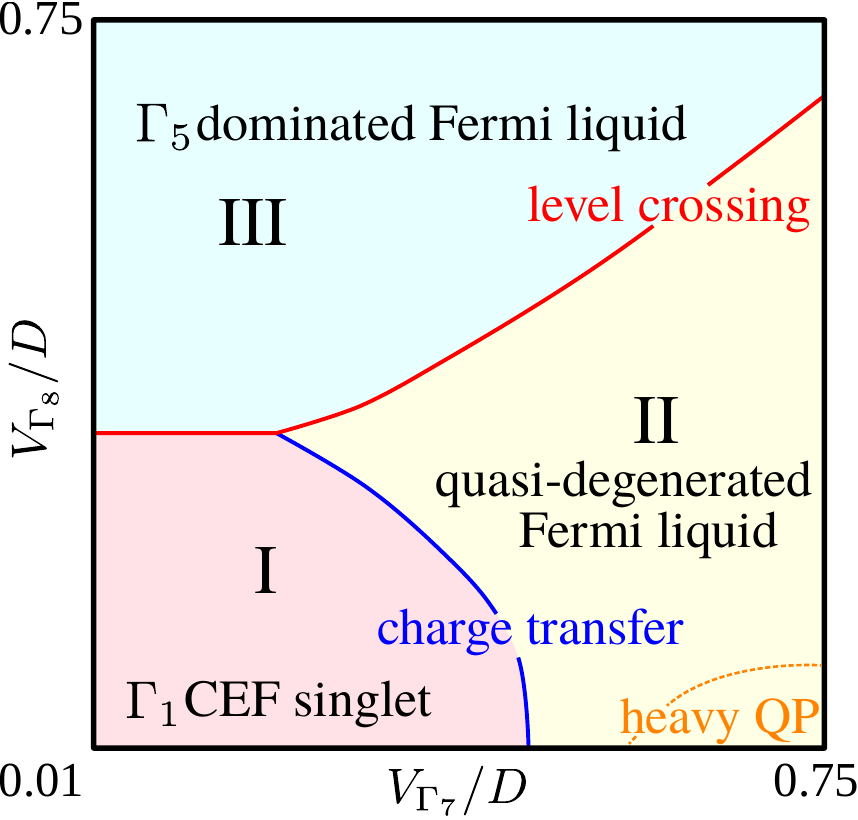}
		\label{Fig:schmatic_phase}
	}
	\subfigure[]{
		\includegraphics[width=0.4\linewidth]{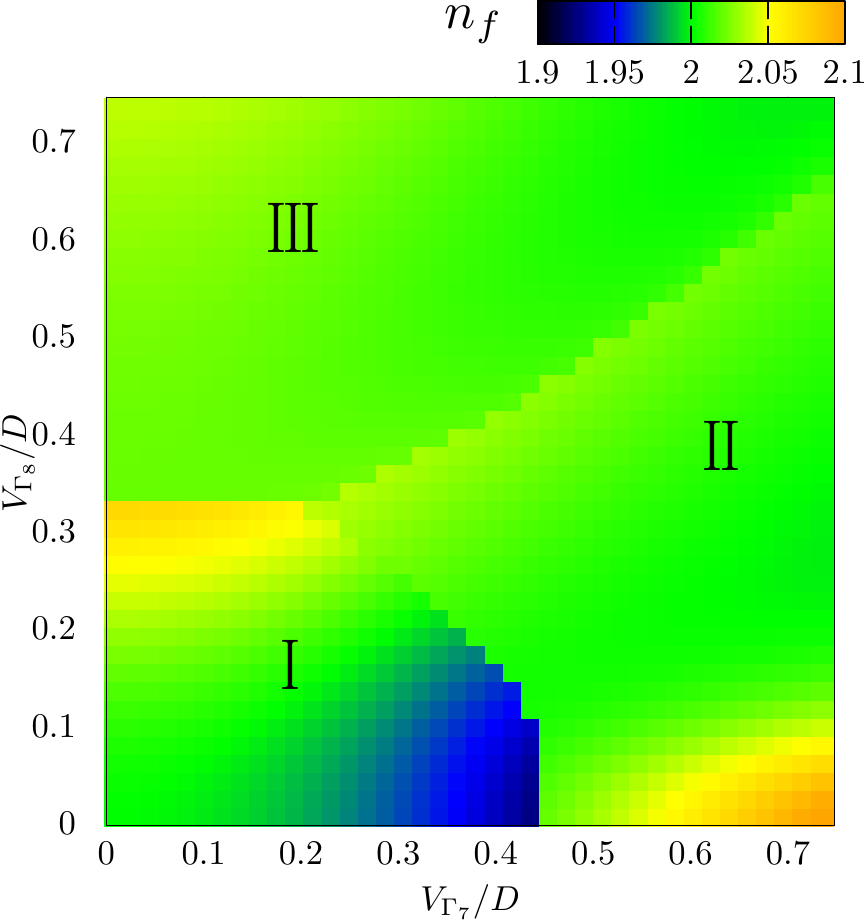}
		\label{Fig:tot_nf}
	}
	\caption{ 
		Results obtained in the RISB SPA for the case of $N=4$, $E_f=-3.3D$, $B_{40}=0.0001D$ and $U=2.0D$ in the $V_{\Gamma_7}$-$V_{\Gamma_8}$ plane:
	(a) Schematic phase diagram in the $V_{\Gamma_7}$-$V_{\Gamma_8}$ plane.
	Phases I, \II, and \III denote the different phases separated by the first-order transitions.
(b) Total number of pseudo-fermions per site $n_f$.}
\end{figure*}

The schematic phase diagram in the $V_{\Gamma_7}$-$V_{\Gamma_8}$ plane is exhibited in Fig. \ref{Fig:cef2}. 
We found three phases, I, \II, and \III, separated by the first-order transitions.
Here, the first-order transitions mean that Eqs.(\ref{eqs1})-(\ref{eqs5}) have stable and metastable states in the vicinity of the phase boundaries.
The phase diagram in Fig. \ref{Fig:schmatic_phase} depicts the stable states.

Figure \ref{Fig:tot_nf} exhibits the number of total $f$-electrons per site $n_f$ and indicates that $n_f$ takes around $2.0$ in the above region.
Thus, the $f^2$-configuration system is realized in most region of the $V_{\Gamma_7}$-$V_{\Gamma_8}$ plane except for part of phases I and \II.
In the following subsections, we first discuss the properties of phases I, \II ($n_f\approx2.0$), \II ($n_f\not\approx2.0$), and \III.
As we discuss later, the physical properties in phase \II with $n_f\not\approx2.0$ region are quite different from those in phase \II with $n_f\approx2.0$ region.
Then, we discuss the properties of phase transitions.

\subsection{Phase I: CEF singlet phase}
\begin{figure*}[t!]
	\centering
	\subfigure[]{
		\includegraphics[width=0.32\linewidth]{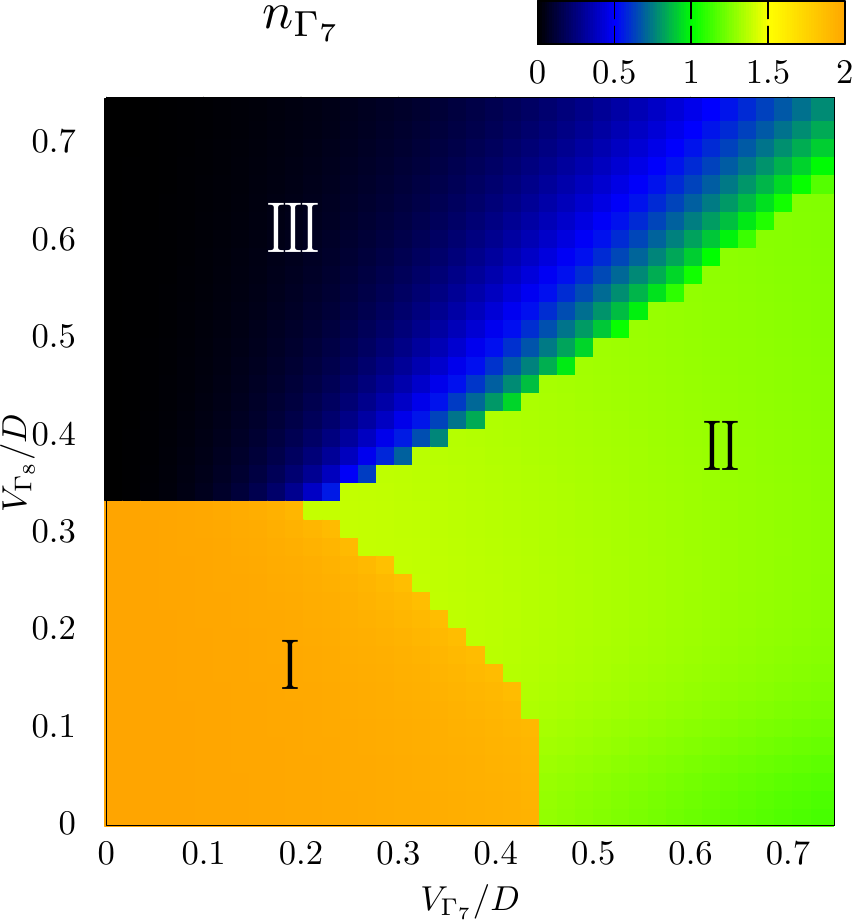}
		\label{Fig:n7}
	}
	\subfigure[]{
		\includegraphics[width=0.32\linewidth]{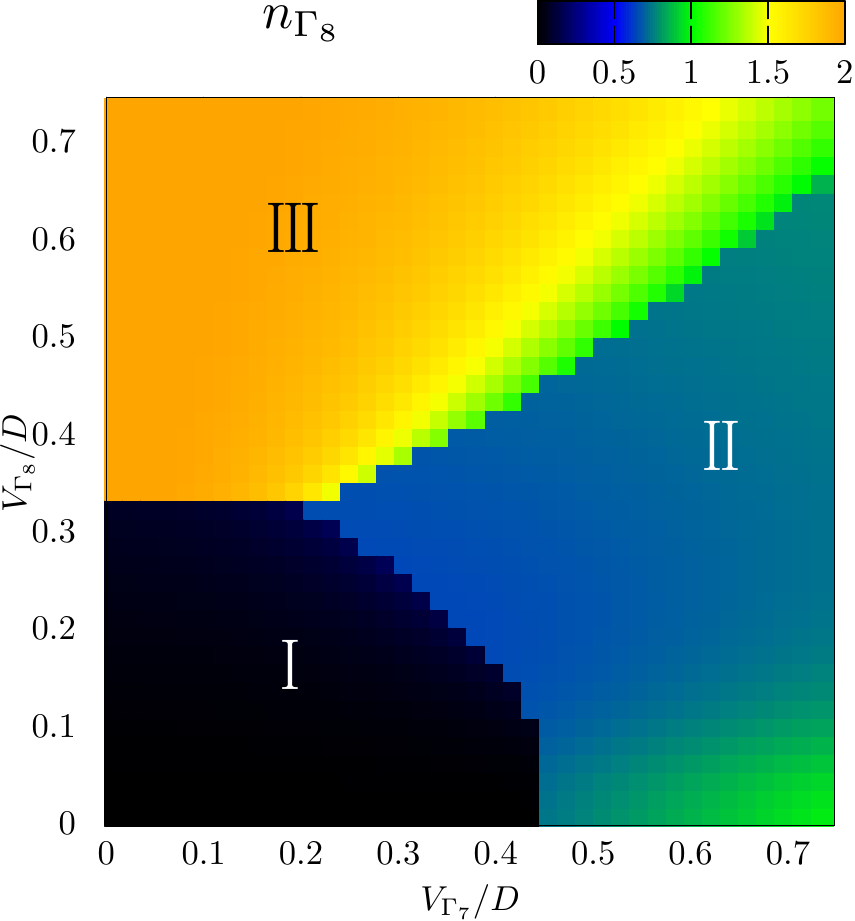}
		\label{Fig:n8}
	}
	\subfigure[]{
		\includegraphics[width=0.32\linewidth]{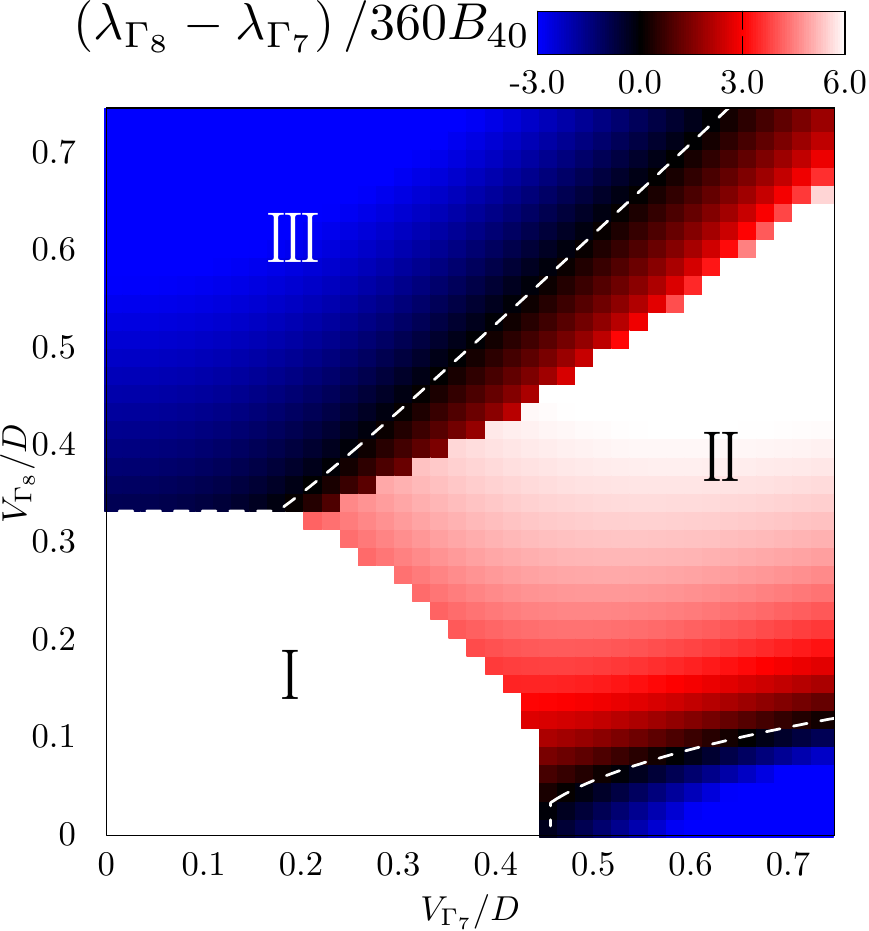}
		\label{Fig:delta_l}
	}
	\subfigure[]{
		\includegraphics[width=0.32\linewidth]{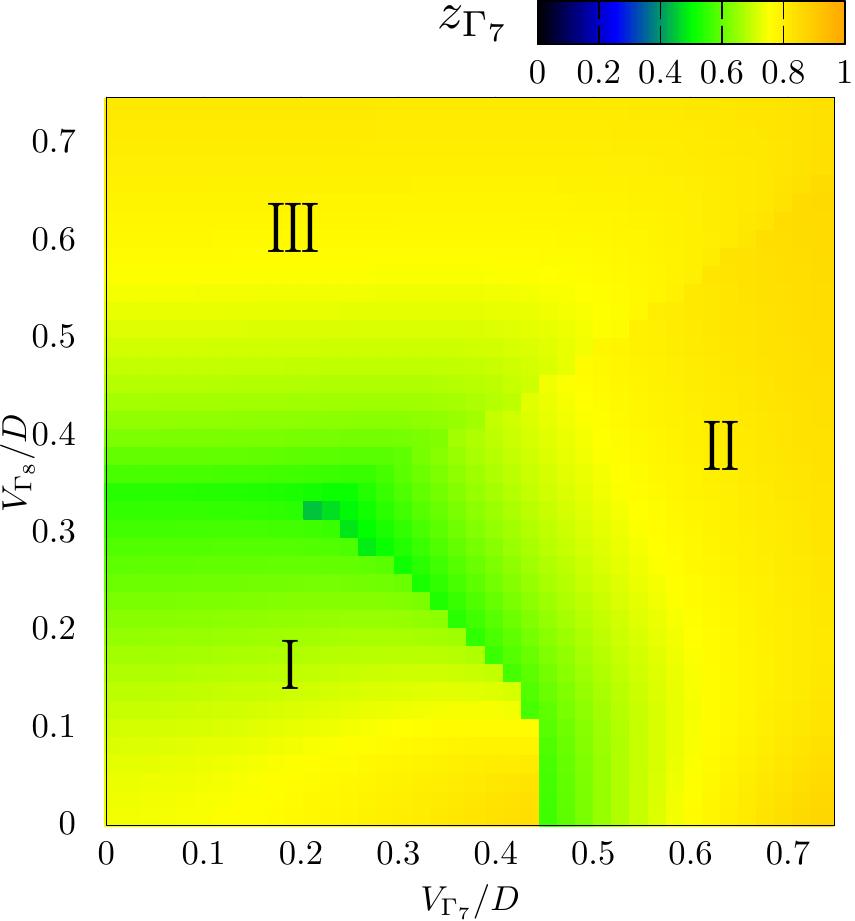}
		\label{Fig:Z7}
	}
	\subfigure[]{
		\includegraphics[width=0.32\linewidth]{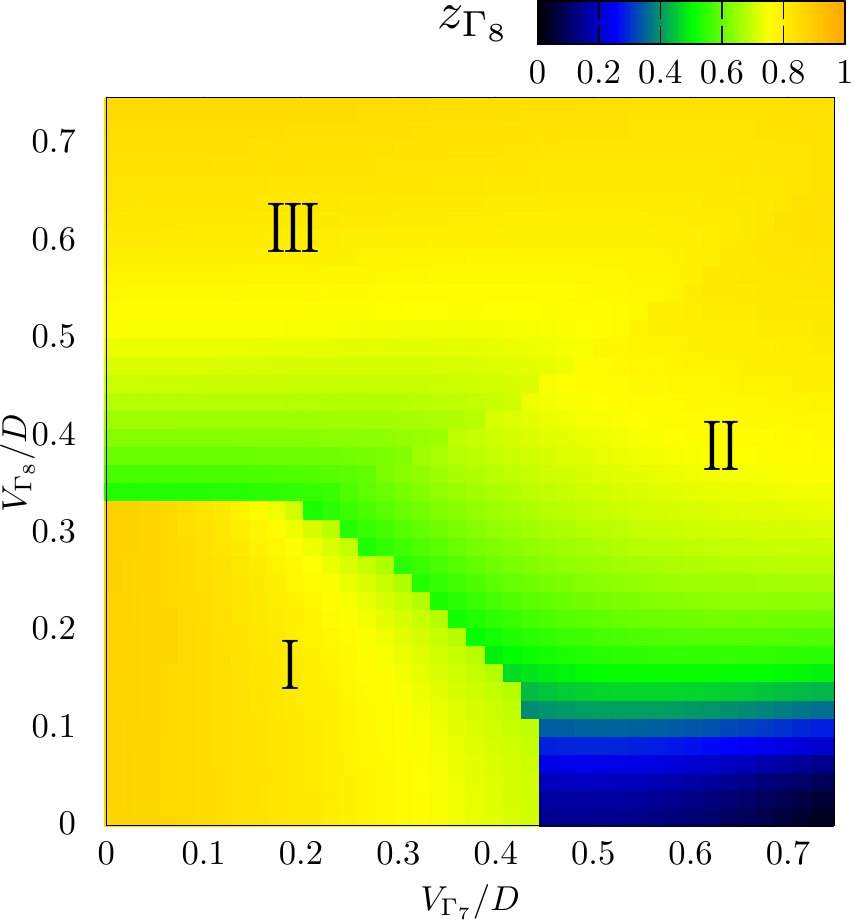}
		\label{Fig:Z8}
	}
	\subfigure[]{
		\includegraphics[width=0.32\linewidth]{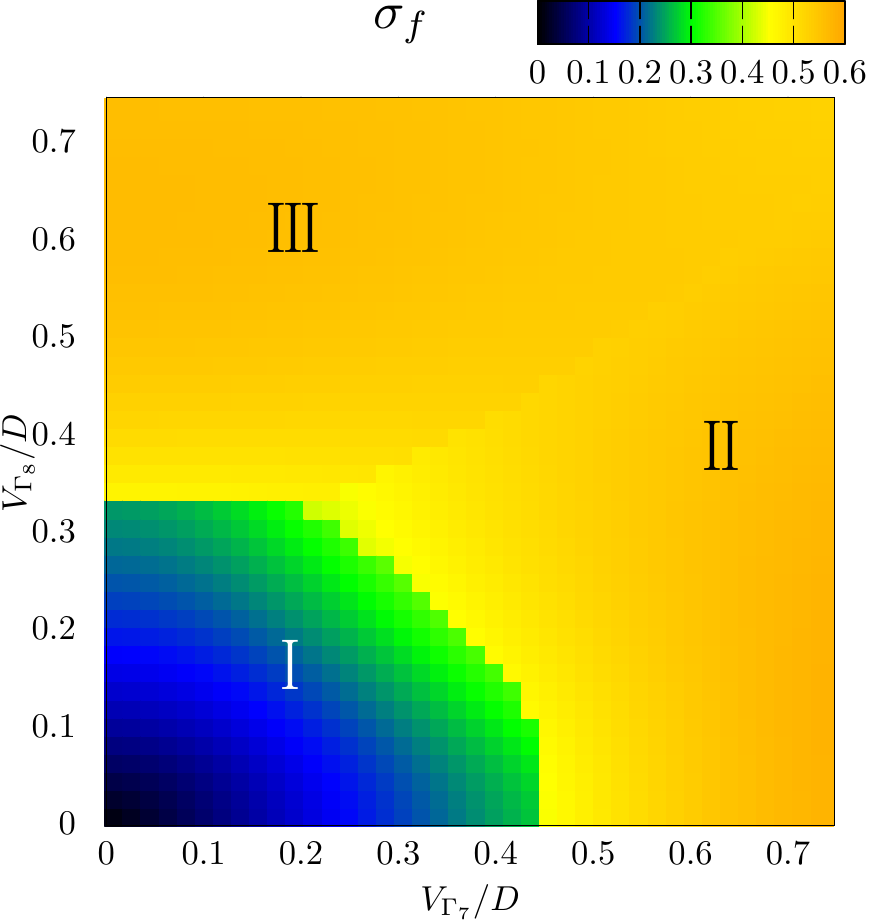}
		\label{Fig:delf2}
	}
	\subfigure[]{
		\includegraphics[width=0.32\linewidth]{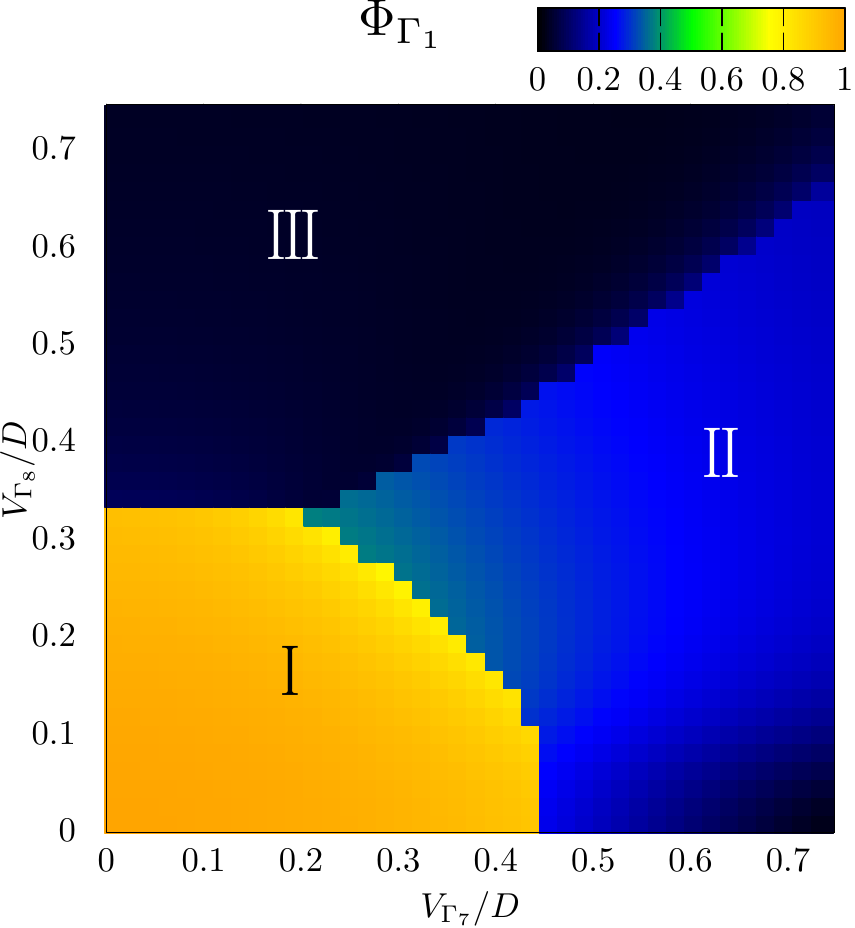}
		\label{Fig:gamma1}
	}
	\subfigure[]{
		\includegraphics[width=0.32\linewidth]{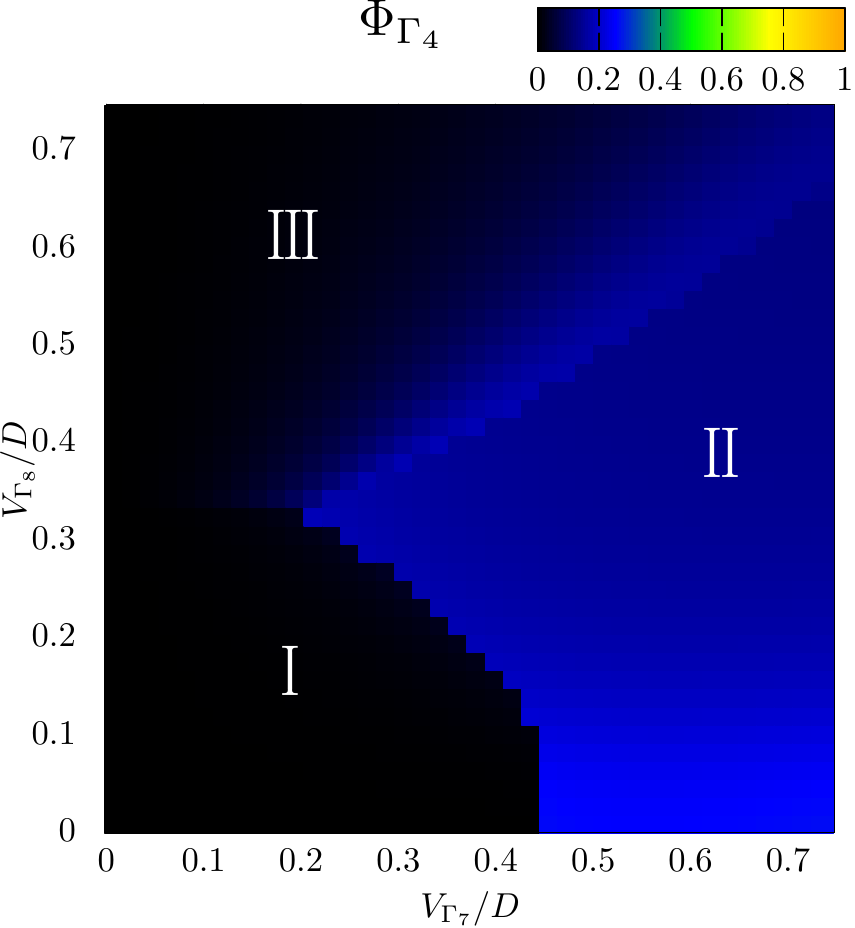}
		\label{Fig:gamma4}
	}
	\subfigure[]{
		\includegraphics[width=0.32\linewidth]{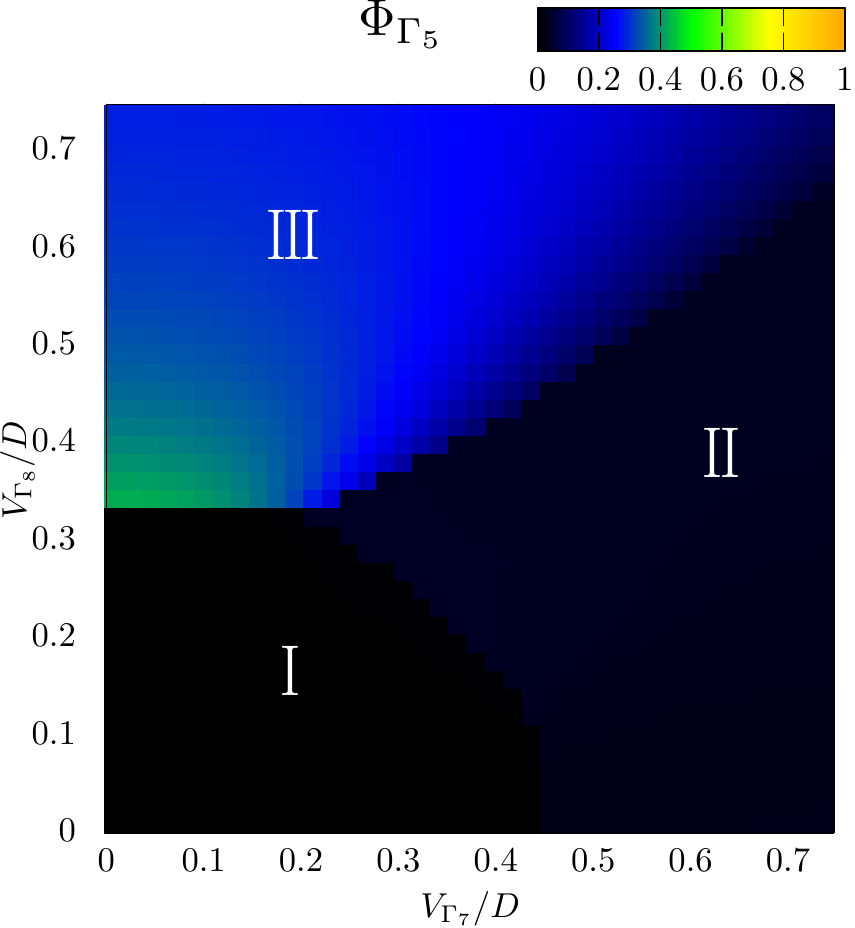}
		\label{Fig:gamma5}
	}
	\caption{ 
		The results obtained in the RISB SPA for the case of $N=4$, $E_f=-3.3D$, $B_{40}=0.0001D$ and $U=2.0D$ in the $V_{\Gamma_7}$-$V_{\Gamma_8}$ plane:
	(a) Pseudo-fermion number per site on the $\Gamma_7$ orbital $n_{\Gamma_7}$.
	(b) Pseudo-fermion number per site on the $\Gamma_8$ orbital $n_{\Gamma_8}$. 
	(c) Effective energy level splitting $(\lambda_{\Gamma_8} - \lambda_{\Gamma_7})$ in the unit of that in the atomic limit $360B_{40} =0.036D$. 
	(d) Renormalization factor of the $\Gamma_7$ orbital $z_{\Gamma_7}$.
	(e) Renormalization factor of the $\Gamma_8$ orbital $z_{\Gamma_8}$. 
	(f) Variance of the $f$-electron number $\sigma_f$.
    (g) Expectation value of the $\Gamma_1$ CEF GS $\Phi_{\Gamma_1}$.
    (h) Expectation value of the $\Gamma_4$ CEF first ES $\Phi_{\Gamma_4}$. 
	(i) Expectation value of the $\Gamma_5$ CEF third ES $\Phi_{\Gamma_5}$.
}
\end{figure*}

Figures \ref{Fig:n7} and \ref{Fig:n8} show the pseudo-fermion number per site on the $\Gamma_7$ orbital $n_{\Gamma_7}$ and that on the $\Gamma_8$ orbital $n_{\Gamma_8}$, respectively.
In phase I, the $\Gamma_7$ orbital is almost doubly occupied by the pseudo fermions while the $\Gamma_8$ orbital is almost empty.
This result indicates that the $\Gamma_7$ ($\Gamma_8$) QP energy level is lower (higher) than the Fermi energy.
Figure \ref{Fig:delta_l} exhibits the energy gap between the two QP energy levels, where the energy gap in the atomic limit ($\Delta_{\mathrm{CEF}}$) is taken as a unit.
The region $\lambda_{\Gamma_8}-\lambda_{\Gamma_7}>6\Delta_{\mathrm{CEF}}$ is displayed in white in this figure.
The results of $n_{\Gamma_7}$, $n_{\Gamma_8}$, and the large energy gap indicate that both QP energy levels are not located in the vicinity of the Fermi level.
In this regard, it is difficult to form the heavy QPs in phase I because of the absence of the QP energy level on the Fermi energy.
Indeed, the renormalization factors $z_{\Gamma_7}$ and $z_{\Gamma_8}$ do not decay toward zero in phase I as shown in Figs. \ref{Fig:Z7} and \ref{Fig:Z8}.

From the viewpoint of the localized $f$-electron nature, or CEF states,
we can conclude that the $f$-electrons are well localized as the $\Gamma_1$ CEF state in phase I.
Figure \ref{Fig:delf2} shows the variance of the $f$-electron number $\sigma_f$.
The expectation values of the CEF states are almost exhausted by the CEF states in the $f^2$-configuration owing to the small amplitude of $\sigma_f$.
In addition, Figs. \ref{Fig:gamma1}-\ref{Fig:gamma5} show the expectation values of the CEF GS $\Phi_{\Gamma_1}$ and the CEF ESs $\Phi_{\Gamma_4}$ and $\Phi_{\Gamma_5}$ in the $f^2$-configuration.
Here, we omit to show the result of $\Phi_{\Gamma_3}$ because it behaves similar to $\Phi_{\Gamma_4}$ in the whole $V_{\Gamma_7}$-$V_{\Gamma_8}$ plane.
According to these figures, the expectation values of CEF states are almost exhausted by the $\Gamma_1$ CEF GS in phase I.
These behaviors are consistent with the \lq\lq $\Gamma_1$ CEF singlet phase" pointed out from the NRG study based on the singlet-triplet model \cite{Hattori2005}.

\subsection{Phase II ($n_f\approx2.0$): quasi-degenerated FL}

In contrast to phase I, the $f$-electrons are well itinerant in phase \II.
In the $n_f\approx 2.0$ region, both $n_{\Gamma_7}$ and $n_{\Gamma_8}$ are almost constant and take non-integer values as seen in Figs. \ref{Fig:n7} and \ref{Fig:n8}, i.e., $n_{\Gamma_7}\approx1.4$ and $n_{\Gamma_8}\approx 0.6$.  
Hence, although the QP energy levels are located near the Fermi energy, the renormalization factors do not decay toward zero as seen in Figs. \ref{Fig:Z7} and \ref{Fig:Z8}.

From the viewpoint of the CEF states, phase \II can be regarded as a quasi-degenerated CEF state.
Figure \ref{Fig:delf2} exhibits that the variance $\sigma_f$ exceeds $0.4$.
The large variance compared with phase I indicates the importance of CEF states in the $f^1$ and $f^3$ configurations.
The amplitudes of $\Phi_{\Gamma}$ in the $f^2$-configuration hold the relation $\Phi_{\Gamma_1} > \Phi_{\Gamma_4} > \Phi_{\Gamma_5}$, as seen in Figs. \ref{Fig:gamma1}-\ref{Fig:gamma5}.
The order of amplitudes $\Phi_{\Gamma}$ corresponds to the CEF energy-level scheme exhibited in Fig. \ref{Fig:cef2}.
Thus, in the $f^2$-configuration, phase \II forms a quasi-degenerated CEF state with the $\Gamma_1$ CEF GS.
Hereafter we call phase \II \lq\lq quasi-degenerated FL".
The properties of this phase are associated with the Kondo-Yosida singlet phase discussed in the previous study~\cite{Nishiyama2013phD}.

\subsection{Phase II ($n_f\not\approx2.0$): heavy QP}

The heavy QP on the $\Gamma_8$ orbital is realized in phase \II with $n_f\not\approx 2.1$ region despite the fact that $n_f$ is not an integer as seen in Figs. \ref{Fig:tot_nf} and \ref{Fig:Z8}.
The heavy QP arises in the intermediate valence because of the properties of phase II, i.e., both $n_{\Gamma_7}$ and $n_{\Gamma_8}$ do not take integer values at the $f^2$-configuration.
Namely, $n_{\Gamma_8}$ approaches $1.0$ by increasing $n_f$ from the $f^2$-configuration, and then the $\Gamma_8$ orbital fulfills the condition to form the heavy QP.

The localized $f$-electron nature is also different from phase \II with $n_f\approx2.0$.
It is remarkable that $\Phi_{\Gamma_1}$ becomes smaller than the $n_f\approx2.0$ region as seen in Fig. \ref{Fig:gamma1}.
This result indicates that the $\Gamma_1$ CEF GS competes with forming the heavy QP.

\begin{figure}[t!]
	\centering
	\includegraphics[width=0.9\linewidth]{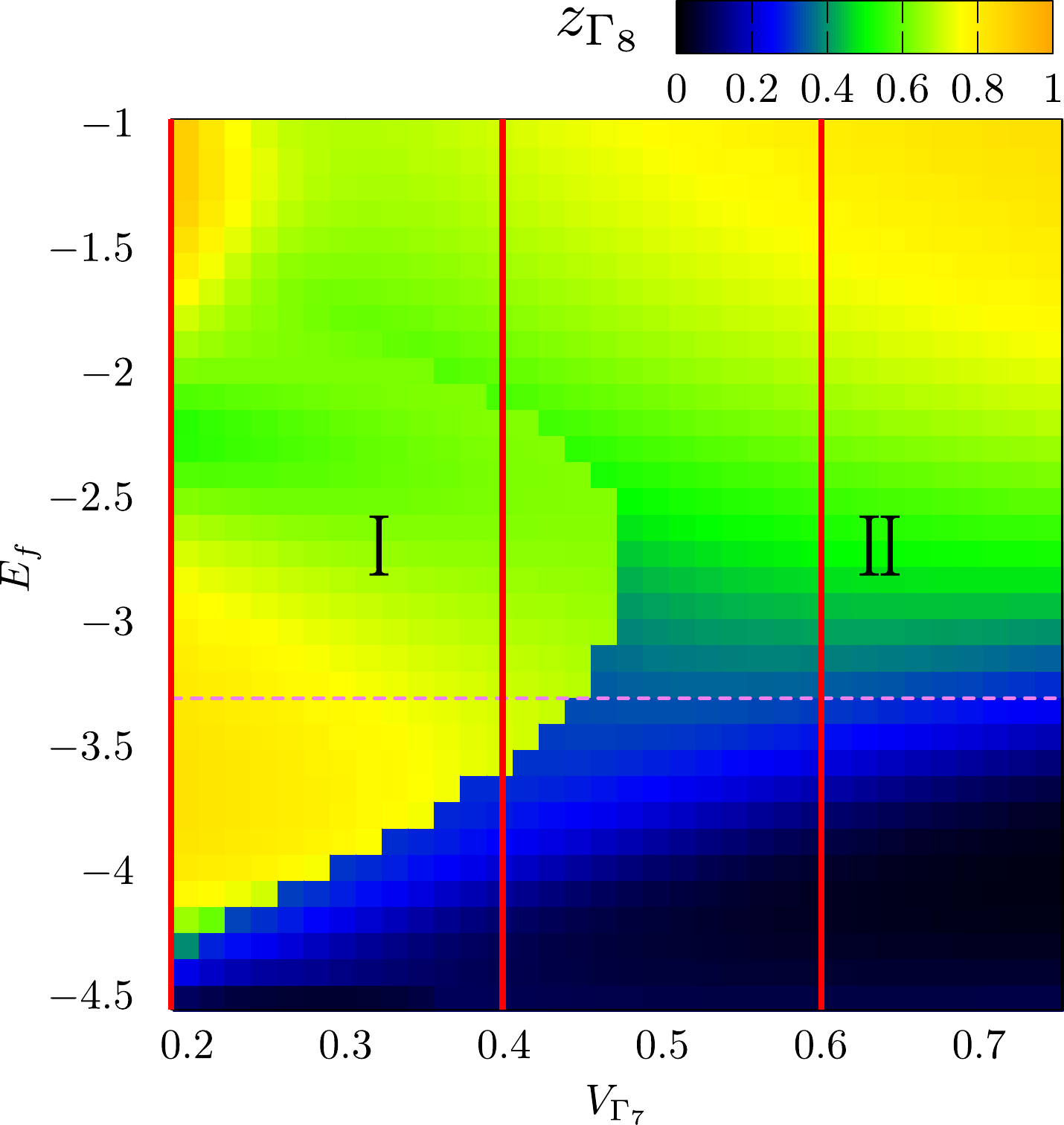}
	\caption{ 
		Renormalization factor on the $\Gamma_8$ orbital as functions of $V_{\Gamma_7}$ and $E_f$.
		Dashed line corresponds to Fig.~4(e) with $V_{\Gamma_8}=0.1D$.
		The three red lines indicate $V_{\Gamma_7}=0,2D$, $0.4D$, and $0.6D$, respectively.}
	\label{Fig:pV7Ef_z8}
\end{figure}

\begin{figure}[t!]
	\centering
	\subfigure[]{
		\includegraphics[width=0.7\linewidth]{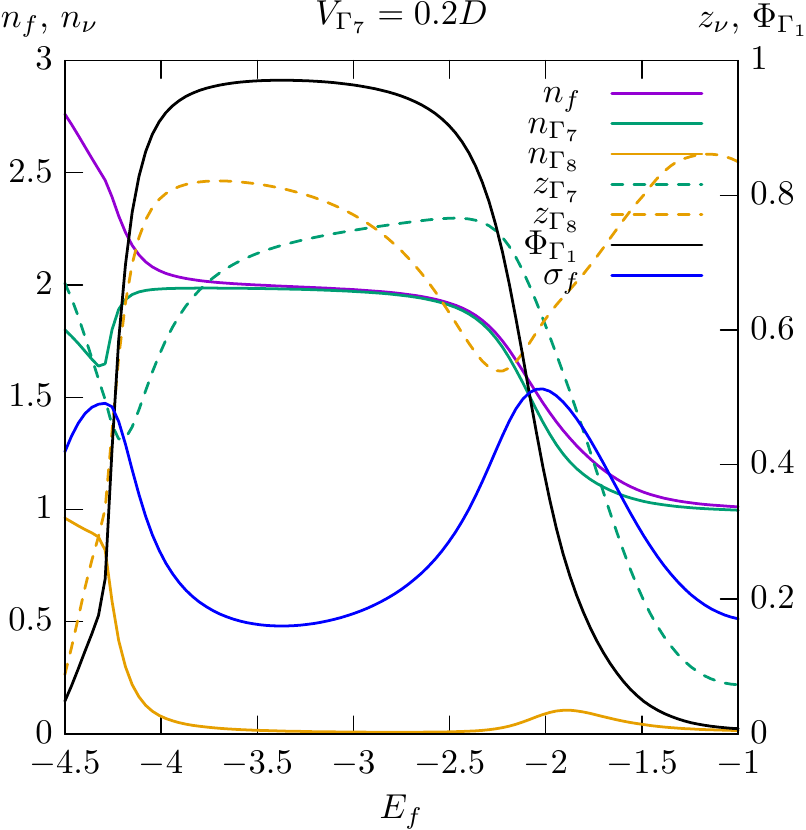}
		\label{Fig:pV7Ef_V02}
	}
	\subfigure[]{
		\includegraphics[width=0.7\linewidth]{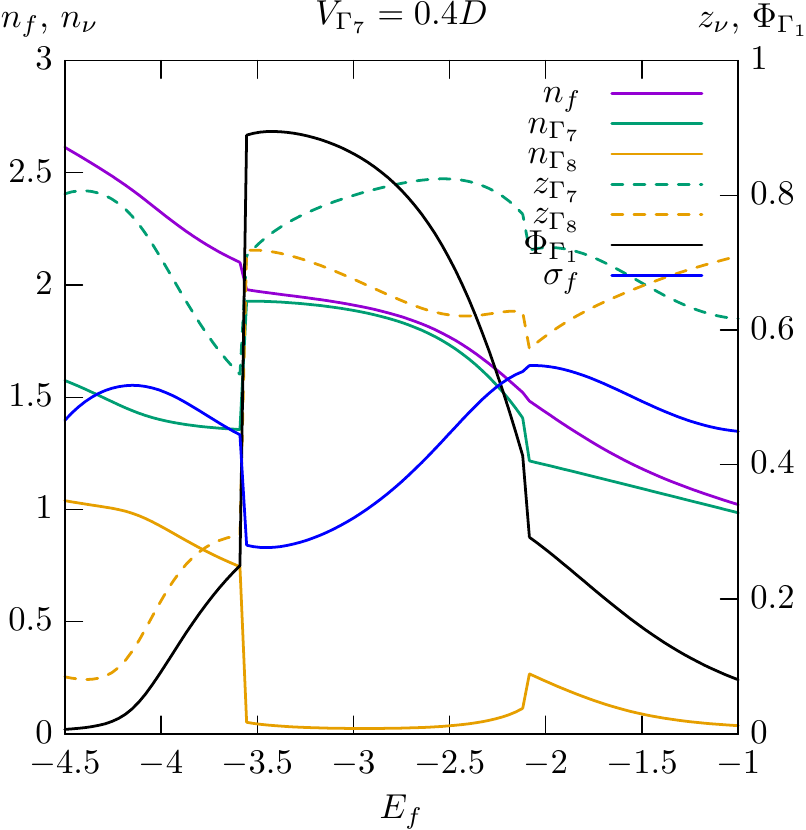}
		\label{Fig:pV7Ef_V04}
	}
	\subfigure[]{
		\includegraphics[width=0.7\linewidth]{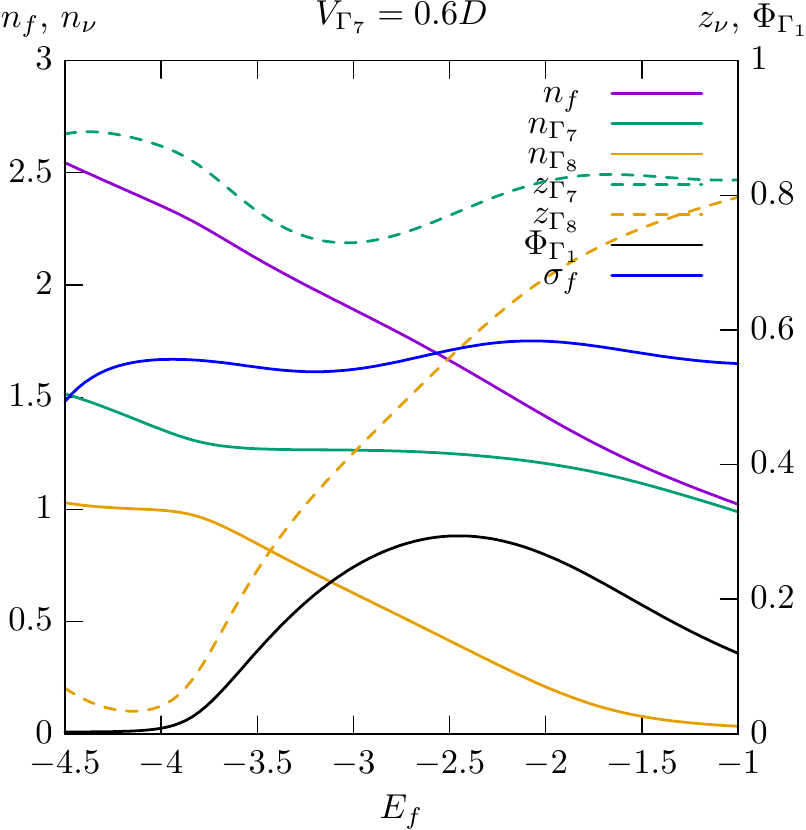}
		\label{Fig:pV7Ef_V06}
	}
	\caption{ 
		Physical properties along (a) $V=0.2D$ (b) $V=0.4D$, and (c) $V=0.6D$ lines in Fig. 5.
		We exhibit the total number of $f$-electrons (violet solid line), the number of pseudo fermions on the $\Gamma_7$ orbital (green solid line) and that on the $\Gamma_8$ orbital (yellow solid line), the renormalization factor on the $\Gamma_7$ orbital (green dashed line) and that on the $\Gamma_8$ orbital (yellow dashed line), and the expectation value of the $\Gamma_1$ CEF state (black solid line).
	}
\end{figure}

According to Fig. \ref{Fig:tot_nf}, the heavy QP arises in the intermediate valence region.
In order to clarify this claim, Fig. \ref{Fig:pV7Ef_z8} exhibits the $E_f$ dependence of $z_{\Gamma_8}$.
Here, $V_{\Gamma_8}$ is fixed as $0.1D$.
The dashed line indicates $E_f=-3.3D$, and the heavy QP robustly exists in phase \II lower than this $f$-electron energy.
Figures \ref{Fig:pV7Ef_V02}, \ref{Fig:pV7Ef_V04}, and \ref{Fig:pV7Ef_V06} exhibit physical properties along $V_{\Gamma_7}=0.2D$, $0.4D$, and $0.6D$ lines, which correspond to the red lines written in Fig. \ref{Fig:pV7Ef_z8}, respectively.
These figures indicate that only the $\Gamma_8$ orbital forms the heavy QP when $n_{\Gamma_8}\approx 1.0$.
In Figs. \ref{Fig:pV7Ef_V04} and \ref{Fig:pV7Ef_V06}, the smallest $z_{\Gamma_8}$ points are located in the intermediate valence region.
In addition, the amplitude of $\Phi_{\Gamma_1}$ are almost zero in these points.
Thus, we conclude that the heavy QP are originated from the intermediate valence region and competes with the $\Gamma_1$ CEF state.


\subsection{Phase \III: $\Gamma_5$ dominated FL}

In phase \III, the $f$-electrons are also itinerant.
However, the occupation numbers of the pseudo-fermions and the distribution of the CEF expectation values are totally different from phase \II.
Figures \ref{Fig:n7} and \ref{Fig:n8} indicate that two pseudo fermions per site are almost completely distributed on the $\Gamma_8$ orbital even though the condition $\varepsilon_{\Gamma_8}>\varepsilon_{\Gamma_7}$.
In contrast to phase I, the $f$-electrons can behave as the itinerant electrons owing to the four-fold degeneracy of the $\Gamma_8$ orbital.

Since there are no $f$-electrons on the $\Gamma_7$ orbital, the only $\Gamma_8$ orbital is possible to form the heavy QP.
However, Fig. \ref{Fig:n8} shows no heavy QP behaviors despite the fact that $n_{\Gamma_8}$ is close to an integer value.
Actually, we have found the region $z_{\Gamma_8}<<1$ in phase \III, although this region is metastable and is covered with phase I.

The $\Gamma_5$ CEF ES shows the largest expectation value in the $f^2$ configuration as seen in Figs. \ref{Fig:gamma1}-\ref{Fig:gamma5}.
In addition, Fig. \ref{Fig:gamma1} shows that the $\Phi_{\Gamma_1}$ is almost zero in this phase.
These results indicate that the CEF energy-level of the $\Gamma_5$ CEF state becomes effectively lower than that of the $\Gamma_1$ CEF state owing to the anisotropy of hybridizations.
In other words, the sign of the \lq\lq effective CEF parameter" $\tilde{B}_{40}$, which characterizes the effective CEF energy-level scheme, changes to a negative value deducing from Fig. \ref{Fig:cef2}.
Thus, we call this phase \lq\lq$\Gamma_5$ dominated FL".
We mention that this phase has not been reported in the previous studies based on the singlet-triplet model \cite{Nishiyama2013phD}.
Instead, the previous studies have reported the $\Gamma_4$ CEF triplet phase in the similar region of phase \III.

\begin{figure}[t!]
	\centering
		\includegraphics[width=0.9\linewidth]{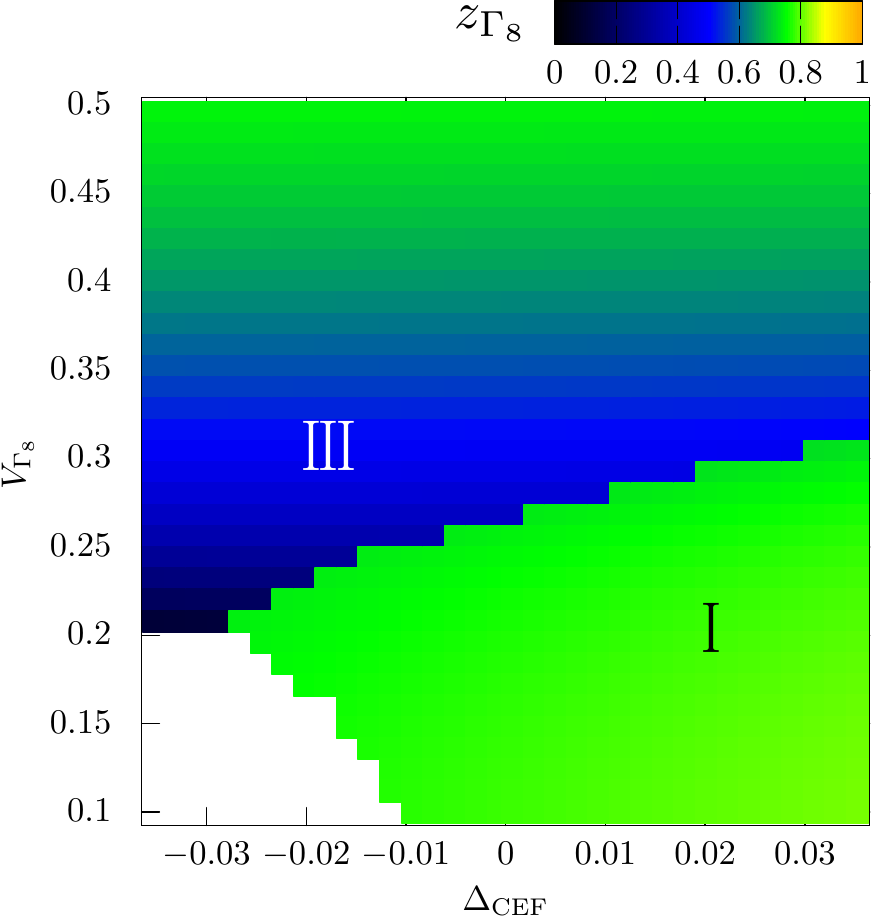}
		\caption{
			Renormalization factor on the $\Gamma_8$ orbital as functions of $V_{\Gamma_8}$ and $\Delta_{\mathrm{CEF}}$.
			Other parameters are fixed as $N=4$, $E_f=-3.0D$, $U=2.0D$, and $V_{\Gamma_7}=0.15D$.
			White region indicates phase \III after the Brinkman-Rice transition.
		}
		\label{Fig:pV8Delta_z8}
\end{figure}

Since the $\Gamma_5$ CEF state is associated with phase \III,
one might think that decreasing $\Delta_{\mathrm{CEF}}$ enlarges the region of phase \III.
In this sense, the heavy QP in phase \III may arise even in the $\Gamma_1$ CEF GS system.
To confirm this point, Fig. \ref{Fig:pV8Delta_z8} exhibits the $\Delta_{\mathrm{CEF}}$ dependence of $z_{\Gamma_8}$ in the case of $V_{\Gamma_7}=0.15D$.
Note that the RISB SPA cannot evaluate phase \III below the Brinkman-Rice transition: $z_{\Gamma_8}=0$.
To depict phase I region, we assume that the free energy of phase \III below the Brinkman-Rice transition is same as that at the transition point with same $\Delta_{\mathrm{CEF}}$.

Figure~\ref{Fig:pV8Delta_z8} indicates that phase I robustly exists even in the $\Gamma_5$ CEF GS system.
This behavior can be understood as follows.
Phase \III exhibits the Brinkman-Rice transition at the finite $V_{\Gamma_8}$.
At this transition point, the free energy for the $f$-electrons will be saturated to the atomic limit.
On the other hand, in phase I, the free energy for the $f$-electrons varies continuously toward the atomic limit.
Therefore, in finite hybridization region, phase I can be stable even in the $\Gamma_5$ CEF GS system.

Figure \ref{Fig:pV8Delta_z8} shows the heavy QP behavior and the Brinkman-Rice transition in phase \III, this region is not the $\Gamma_1$ CEF GS system
Thus, we conclude that the robustly existing phase I prevents forming the heavy QP in phase \III with the $\Gamma_1$ CEF GS system.

\subsection{Phase transition between phases I and \II}

Let us discuss the properties of phase transitions.
The transition between phases I and \II can be regarded as the charge-transfer transition accompanied with the change of $n_{\Gamma_7}$ and $n_{\Gamma_8}$.
This transition has been also proposed in the singlet-triplet model~\cite{Shiina2005},
and is analogous to the valence transition in heavy electron systems~\cite{Onishi2000,Watanabe2006a,Watanabe2007,Watanabe2009}.
Watanabe et.al. have pointed out that the first-order valence transition occurs when the Coulomb interaction $U_{fc}$ between the $f$-electrons and conduction electrons
is strong enough in the extended single-orbital periodic Anderson model~\cite{Watanabe2006a,Watanabe2007}.
In the present model, the inter-orbital interaction between the $\Gamma_7$ orbital and $\Gamma_8$ orbitals plays a similar role of $U_{fc}$.

Present system exhibits the quantum critical end points (QCEP) in the both sides of the $f^2$-configuration:
$V_{\Gamma_7}=0.3D$ and $E_f=-2.0D$ in the $f^1$-$f^2$ intermediate valence region, and $V_{\Gamma_7}=0.21D$ and $E_f=-4.25D$ in the $f^2$-$f^3$ intermediate valence region.
Indeed, while Fig. \ref{Fig:pV7Ef_V04} shows the first-order phase transitions, both the transitions change to crossover behaviors in Fig. \ref{Fig:pV7Ef_V02}.
The total number of $f$-electrons $n_f$ are 1.6 and 2.4 at the QCEPs, respectively.
Both points are located far from the $f^2$-configuration.

\subsection{phase transition between phase \III and others}

The phase transition between phases \III and others is concluded as the CEF energy-level crossing because of the physical properties of phase \III.
Owing to the large amplitude of $V_{\Gamma_8}$, the $f$-electrons favor to be located on the $\Gamma_8$ orbital because of the gain in the kinetic energy even in the case of $\varepsilon_{\Gamma_8} > \varepsilon_{\Gamma_7}$.
Namely, the $\Gamma_8$ CEF state in the $f^1$-configuration becomes effectively lower than the $\Gamma_7$ CEF state, and thus the effective CEF parameter $\tilde{B_{40}}$ becomes negative.
In this situation, phase \III is naturally induced as seen in Fig. \ref{Fig:cef2}.

\section{Discussion}

\subsection{Difference from singlet-triplet model}

The phase diagram [Fig. \ref{Fig:schmatic_phase}] exhibits different properties from that evaluated by the singlet-triplet model studied in Refs. \citen{Hattori2005,Nishiyama2013phD}.
First, in the $V_{\Gamma_7}<V_{\Gamma_8}$ region, the previous studies suggested the $\Gamma_4$ triplet CEF phase, although we found the new phase, i.e., the $\Gamma_5$ dominated FL.
This difference implies that the $\Gamma_5$ CEF state cannot be ignored to investigate the $V_{\Gamma_7}<V_{\Gamma_8}$ region.

Second, in the $V_{\Gamma_7}>V_{\Gamma_8}$ region, the present study indicates the quasi-degenerated FL, while the previous study pointed out that the $\Gamma_1$ CEF singlet state robustly survives.
This difference depends on whether CEF states in the $f^3$-configuration is taken into account or not.
As we discussed in Sect. 4, phase \II (including $n_f\not\approx2$ region) cannot be explained by specific CEF states in the $f^2$ configuration because of the large amplitude of $\sigma_f$.
Namely, the CEF states in the $f^3$ configuration play an important role in this phase.
On the other hand, in the singlet-triplet model, CEF states in the $f^3$-configuration are omitted, and thus the $\Gamma_1$ singlet CEF phase robustly survives.

Third, the previous study has pointed out the existence of the NFL nearby the phase transition between the $\Gamma_4$ triplet CEF phase and others.
It is suggested that the mass enhancement behavior observed in $\mathrm{UBe_{13}}$ can be understood by this NFL.
The present study, however, cannot capture this behavior owing to the SPA.
We mention that the properties of the NFL in between the phases I and \III may change from those reported in the singlet-triplet model because of the $\Gamma_5$ dominated FL.
Further studies beyond the SPA are required to clarify this point.
Although Ref. \citen{shimizu2005a} has already studied the impurity system taking into account all the CEF states by using the NRG, the existence of the NFL has not been discussed.
Thus, it is also worth revisiting the impurity system.

\subsection{Relation between CEF states and QP configurations}

In the case of $n_f\approx2.0$ and $\sigma_f<<1$, expectation values of CEF states are exhausted by specific CEF states.
Then, an $f$-electron configuration of a QP should be associated in some way with that of the dominant CEF states.
In the present system, we found two regions satisfying the conditions: the $\Gamma_1$ CEF singlet phase and the heavy QP in phase \III with the $\Gamma_5$ CEF GS.
Here, although we do not present $\sigma_f$ behavior in the heavy QP in phase \III, it is obvious that the Brinkman-Rice transition induces $\sigma_f=0.0$.

According to Table \ref{tab:1}, the QP configuration, $n_{\Gamma_7}$ and $n_{\Gamma_8}$, are associated with the dominant Fock state term of the effective CEF GS.
As an example, the QP configuration in phase I is $n_{\Gamma_7}\approx 2.0$ and $n_{\Gamma_8}\approx 0.0$, while the dominant Fock state in the $\Gamma_1$ CEF state is $\left\{ +\Gamma_7, -\Gamma_7\right\}$.
It is obvious that both the $f$-electron configurations are same.
Such the relation is also found in the case of the $\Gamma_5$ CEF GS and the heavy QP configuration.
Other Fock states composing the CEF GS are irrelevant to the QP configuration.
Moreover, in accordance with this relation, we can also understand why the $n_{\Gamma_7}\approx 1.0$ and $n_{\Gamma_8}\approx 1.0$ situation is absent in the present phase diagram.
Namely, the $\Gamma_3$ and $\Gamma_4$ CEF states, in which a dominant Fock state induce such the QP configuration, cannot be the CEF GS in the present system as seen in Fig. \ref{Fig:cef2}.
However, since the present model ignores the inter-orbital hybridizations, it is intriguing problem whether the above relation is kept even in the realistic model.

Previous studies have been proposed that the QP configuration is same as the CEF configuration by using the Kotliar-Ruckenstein slave boson (KRSB) formalism~\cite{ikeda1997,Kusunose2005}.
Although these results are inconsistent with the present result, Ref.~\citen{Lechermann2007} has pointed out that the KRSB formalism cannot be applied to a multi-orbital periodic Anderson model with non-density-density type interactions.

\subsection{Mass enhancement mechanism}

\begin{figure}[t!]
	\centering
		\includegraphics[width=0.9\linewidth]{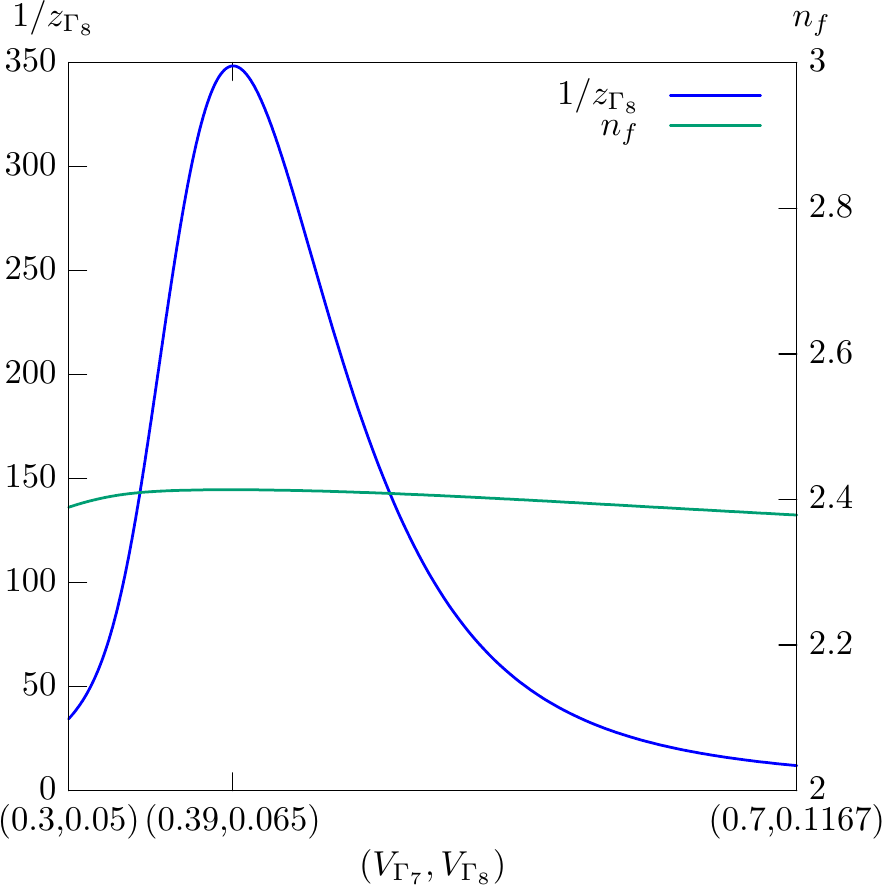}
		\caption{Inverse of the renormalization factor $z_{\Gamma_8}$ (blue solid line), which can be roughly estimated as a mass enhancement factor, as a function of hybridizations.
		The number of total $f$-electrons (green solid line) is also exhibited.
		Parameters are fixed as $N=4.0$, $E_f=-4.1D$, $U=2.0D$, and $B_{40}=0.0001D$.
	}
		\label{Fig:UBe13}
\end{figure}

By using the RISB SPA, we found the heavy QP in the intermediate valence region.
Owing to the orbital degree of freedom on the $\Gamma_8$ orbital, arising NFL behavior is expected in this region.
This region may provide us new insights into the effective mass enhancement mechanism observed in $\mathrm{UBe_{13}}$.
In fact, we can reproduce the peak structure of the effective mass enhancement as a function of the lattice constant~\cite{Kim1990}.
It is expected that the lattice constant affects the hybridizations.
When we assume the anisotropic hybridizations, $V_{\Gamma_7} =6V_{\Gamma_8}$,
the inverse of the renormalization factor $1/z_{\Gamma_8}$, which can be roughly estimated as the mass enhancement factor, exhibits a peak structure as seen in Fig. \ref{Fig:UBe13}.
However, the present results cannot reproduce the discontinuity of the peak structure observed in Ref.~\citen{Kim1990}.
Since the present study cannot discuss the NFL behavior, further studies going beyond the SPA is required.

On the other hand, our results hard to explain the mass enhancement mechanism observed in $\mathrm{PrOs_4Sb_{12}}$.
In this compound, it is expected that the $f$-electrons are well localized and forming the quasi-quartet CEF state~\cite{Otsuki2005}.
However, the heavy QP found in the present study competes with the $\Gamma_1$ CEF GS.
In order to realize the heavy QP associated with the quasi-quartet CEF state, introducing inter-orbital hybridizations and/or going beyond the SPA may be important.

\section{Conclusion}

We investigated the $\Gamma_1$ singlet CEF GS system in cubic symmetry by using the RISB SPA.
We found three phases: the CEF singlet phase, the $\Gamma_5$ dominated FL, and the quasi-degenerated FL.
The $\Gamma_5$ dominated FL shows quite different properties compared with the CEF triplet phase found in the singlet-triplet model.

We also discussed the origin of the phase transitions among these phases:
the transition between phase I and \II is regarded as the charge transfer transition,
while the transition between phase \III and other two phases is regarded as the effective CEF energy-level crossing.

We found the heavy QP in the intermediate valence region.
This heavy QP may provide us new insights into the mass enhancement mechanism in $\mathrm{UBe_{13}}$.
On the other hand, at the $f^2$-configuration, we conclude that it is hard to realize the heavy QP unless introducing a realistic model and/or evaluating beyond the SPA.

\section*{Acknowledgement}

We are grateful to Prof. Kazumasa Miyake, Prof. Masao Ogata, Atsushi Tsuruta, Hideaki Maebashi, and Hiroyasu Matsuura for enlightening discussions and useful comments.

\bibliography{main.bbl}
\end{document}